\documentclass[fleqn,usenatbib]{mnras}

\usepackage[T1]{fontenc}
\usepackage{ae,aecompl}

\usepackage{graphicx}   
\usepackage{amsmath}    
\usepackage{amssymb}    

\title[Refinement of pulsar magnetic fields]{Refinement of the timing-based estimator of pulsar magnetic fields}

\author[Biryukov, Astashenok \& Beskin]{
Anton Biryukov$^{1,2}$\thanks{E-mail: ant.biryukov@gmail.com},
Artyom Astashenok$^{3}$\thanks{E-mail:
artyom.astashenok@gmail.com} and
Gregory Beskin$^{4,2}$
\\
$^{1}$Sternberg Astronomical Institute of Lomonosov Moscow State University, 13 Universitetsky pr., Moscow 119234, Russia \\
$^{2}$Kazan Federal University, 18 Kremlyovskaya str., Kazan 420008, Russia \\
$^{3}$Physics, Mathematics \& IT Department, I. Kant Baltic Federal University, 14 A. Nevskogo str., 236041, Kaliningrad, Russia\\
$^{4}$Special Astrophysical Observatory of the Russian Academy of Sciences, Nizhnii Arkhyz, 369167, Russia }

\date{Accepted XXX. Received YYY; in original form ZZZ}

\pubyear{2017}

\begin{document}
\label{firstpage}
\pagerange{\pageref{firstpage}--\pageref{lastpage}}
\maketitle

\begin{abstract}
Numerical simulations of realistic non-vacuum magnetospheres of isolated neutron stars have
shown that pulsar spin-down luminosities depend weakly on the magnetic obliquity $\alpha$. In
particular, $L \propto B^2(1 + \sin^2\alpha)$, where $B$ is the magnetic field strength at the star surface. Being
the most accurate expression to date, this result provides the opportunity to estimate $B$ for a
given radiopulsar with quite a high accuracy. In the current work, we present a refinement of the
classical `magneto-dipolar' formula for pulsar magnetic fields $B_{\rm md} = (3.2\times 10^{19}\mbox{ G})\sqrt{P\dot P}$,
where $P$ is the neutron star spin period. The new, robust timing-based estimator is introduced as
$\log B = \log B_{\rm md} + \Delta_{\rm B}(M, \alpha)$, where the correction $\Delta_{\rm B}$ depends on the equation of state (EOS)
of dense matter, the individual pulsar obliquity $\alpha$ and the mass $M$. Adopting state-of-the-art
statistics for $M$ and $\alpha$ we calculate the distributions of $\Delta_{\rm B}$ for a representative subset of 22 EOSs
that do not contradict observations. It has been found that $\Delta_{\rm B}$ is distributed nearly normally,
with the average in the range $-0.5$ to $-0.25$ dex and standard deviation $\sigma[\Delta_{\rm B}] \approx$ 0.06 to 0.09 dex, depending on the adopted EOS. The latter quantity represents a formal uncertainty
of the corrected estimation of $\log B$ because $\Delta_{\rm B}$ is weakly correlated with $\log B_{\rm md}$. At the
same time, if it is assumed that every considered EOS has the same chance of occurring in
nature, then another, more generalized, estimator $B^* \approx 3B_{\rm md}/7$ can be introduced providing
an unbiased value of the pulsar surface magnetic field with $\sim$30 per cent uncertainty with
68 per cent confidence. Finally, we discuss the possible impact of pulsar timing irregularities
on the timing-based estimation of $B$ and review the astrophysical applications of the obtained
results.

\end{abstract}




\section{Introduction}

Radiopulsars are strongly magnetized neutron stars (NSs). The
large-scale magnetic field, originating in their crusts (and, probably,
in the cores), plays a crucial role in many phenomena related to these
compact remnants of stellar evolution. However, only a few dozen
nearly direct measurements of NS magnetic fields have been made
so far -- typically for high-energy sources in binary systems. These
measurements are based on accretion and/or emission physics, such
as the detection of a cyclotron resonant scattering feature in the
X-ray spectra of a compact object \citep[see for instance][and references therein]{anna12, tien13, revmer15, borg15}.

The magnetic fields of isolated radiopulsars can be routinely
inferred from their timing. Indeed, pulsar rotational histories are to
a large extent driven by the properties of their magnetospheres --
through the charged particle winds and the surface currents acting
on the star. On the other hand, the evolution of the pulsar spin period
P(t) can be directly traced because of the anisotropy of the radio
emission.

The topology of the magnetic field of an isolated NS is typically
assumed to be purely (or predominantly) dipolar, because higher-order multipoles decrease rapidly with distance from the star. In this
case, the braking torque component aligned with the star spin axis
can be calculated as
\begin{equation}
	N =  - \mu^2  \left (\dfrac{\Omega}{c} \right )^3  f(\alpha),
	\label{eq:torque_general}
\end{equation}
where $\mu$ is the dipolar magnetic moment, $\Omega = 2\upi/P$ is the spin
frequency, $c$ is the speed of light, while $f(\alpha)$  is the dimensionless function of the angle $\alpha$ between the magnetic and spin axes.
The cubic dependence of $N$ on the inverse light-cylinder radius
$r_{\rm _{LC}} = c/\Omega$ (the natural size of the magnetosphere) and the quadratic
dependence on the moment $\mu$ probably reflects the dipolar structure
of the field and is common in most theories of pulsar spin-down.\footnote{See, however, the recent paper by \cite{petr16}, in which an alternative
pulsar braking torque $N' \approx N\cdot (\Omega R/c)^{-2}$ was derived for an aligned rotator
(where $R$ is the NS radius), and another way of refining the magnetic field
formula was discussed. Nevertheless, in our research we follow the classical
approach, assuming $N \propto -\mu^2/r^3_{\rm _{LC}}$.} In contrast, the particular form of $f(\alpha)$ is highly dependent on the
adopted physics and can be generically written as
\begin{equation}
	f(\alpha) = x_{\perp} \sin^2\alpha + x_{\parallel}\cos^2\alpha,
	\label{eq:f_general}
\end{equation}
where $x_{\perp}$ and $x_{\parallel}$ are the coefficients describing the losses of
rotational energy by perfectly orthogonal and aligned rotators,
respectively.

For a spherical NS with moment of inertia $I$ and radius $R$, equation (\ref{eq:torque_general}) can be rewritten as the spin-down law
\begin{equation}
    P\dfrac{{\rm d}P}{{\rm d}t} = \dfrac{4\upi^2 R^6}{Ic^3} B^2 f(\alpha),
    \label{eq:spindown_general}
\end{equation}
where $B = \mu/R^3$ is the field strength at the {\it magnetic equator}. This
law explicitly defines $B$ as a function of period $P$ and its derivative
$\dot P$, which is the basis of the timing-based estimation of the magnetic
field. Of course, the precise calculation of $B$ requires the radius $R$,
the moment of inertia $I$ and the magnetic angle $\alpha$ of a given pulsar to
be known as well. The latter quantity is the most crucial -- indeed,
if $x_{\perp} \gg x_{\parallel}$ or vice versa, then even a moderate uncertainty in $\alpha$
will lead to a significant error in the derived value of $B$. Moreover,
when $\alpha$ is completely unknown, only a lower limit of the surface
field can be obtained. On the other hand, the radii and moments of
inertia are expected to be relatively narrowly distributed over the
pulsar population, mostly owing to the plateau in the mass--radius
relationship predicted by realistic equations of state (see Fig.~\ref{fig:eos_list}).
Thus, a confident knowledge of $f(\alpha)$ seems to be the key component
in the calculation of pulsar magnetic fields

Many models of radiopulsar spin-down (in terms of $f(\alpha)$
parametrization) have been discussed in the literature for the last
half-century (see \cite{bes13} for a review).
For instance, the classical but naive model of vacuum `magnetodipolar' losses owing to the radiation-reaction torque assumes
$x_{\perp} \in 0.3-1.3$ and $x_{\parallel} \equiv 0$ \citep{deu55, dg70, good85, melatos2000}. The most frequently used version of this model, with $x_{\perp} = 2/3$, $\alpha = 90\degr$, moment of inertia
$I = I_0 = 10^{45}$ g cm$^2$ and radius $R = R_0 = 10$ km, provides the
`standard' widely used equation for the pulsar magnetic field \citep[e.g.][]{mt77, st83}:
\begin{equation}
    B_{\rm md}(P, \dot P) = \sqrt{\dfrac{3 I_0 c^3}{8\upi^2 R_0^6} \cdot P \dot P} = 3.2\times 10^{19} \sqrt{P \dot P}\mbox{ G}
    \label{eq:B0}
\end{equation}
Even if one disregards the fact that equation (\ref{eq:B0}) is based on unrealistic assumptions (namely, a vacuum NS magnetosphere and
common values of $I$ and $R$ for all pulsars), the obliquity $\alpha =
90\degr$ assumed in $B_{\rm md}$  makes its value formally only a lower limit of the
NS surface field strength.

The `magneto-dipolar' losses vanish for a perfectly aligned rotator ($\alpha
= 0\degr$). On the other hand, earlier investigations of axisymmetric plasma-filled magnetospheres have shown that a NS
with zero obliquity still efficiently loses its rotational energy \citep[e.g][]{gj69, michel73, rs75}. Later, \cite{bes93} obtained
a general analytic solution for a non-vacuum NS magnetosphere with arbitrary $\alpha$, adopting a number of reasonable simplifications.
They found that $x_{\parallel} \approx
x_{\perp} (\Omega R/c)^{-1}$. In this case, the first term of (\ref{eq:f_general})
appears to be almost suppressed because $\Omega R/c \sim 10^{-4}$ to $10^{-2}$ for
actual pulsars \citep
[see also][]{bes07}. Hence, even in
this more physically justified model, the magnetic field estimation
(derived from the corresponding spin-down law $P\dot P \propto \Omega^3 \cos^2\alpha$)
still crucially depends on the obliquity $\alpha$.

It has been proposed in the last decade, however, that the observed spin-down of actual pulsars can be satisfactorily explained
by assuming that the two terms of $f(\alpha)$ contribute comparably to the
braking torque; that is, $x_{\perp} \sim x_{\parallel}$ \citep{xu01, cs06, bars09, bars10, kou15, kou16, ou16}. If this is the case, it can be expected that the 
spindown rate ${\rm d}P/{\rm d}t$ depends relatively weakly on the obliquity $\alpha$, thus
eliminating the problem of `uncertain magnetic field estimation'
described above.

This proposition has been well supported by the results from
direct three-dimensional magnetohydrodynamical (MHD) and
particle-in-cell numerical simulations of oblique pulsar magnetospheres performed by a number of groups \citep{spitkovsky06, kala09, petri12, tche13,
phil14}. It has
been found that $f(\alpha)$ can be approximated by the simple analytic
formula
\begin{equation}
    f(\alpha) = k_0 + k_1\sin^2\alpha,
    \label{eq:spitkovsky_obliq}
\end{equation}
where both $k_0$ and $k_1$ are close to unity and constant to within the
terms proportional to $(\Omega R/c)^2$. In terms of equation (\ref{eq:f_general}), it means
that $x_{\perp} = k_0 + k_1$ and $x_{\parallel} = k_0$ respectively. Being the most accurate
solution obtained so far, this improved and quite simple form of
$f(\alpha)$ provides a way to measure the surface magnetic field of observed isolated pulsars with a relatively high precision, even when
$\alpha$ remains unknown. \footnote{Because it is the result of purely numerical calculations, equation (\ref{eq:spitkovsky_obliq}) has
no transparent physical interpretation at present. However, it can be understood in terms of the so-called symmetric $i_{\rm s}$ and antisymmetric $i_{\rm a}$ 
electric currents. These currents flow within the NS polar caps and are responsible
for the braking torque components that are parallel and perpendicular to
the NS magnetic moment, respectively. Normalizing $i_{\rm s}$ and $i_{\rm a}$ to the local
Goldreich--Julian current $j_{\rm _{GJ}} \approx (\mathbf{\Omega} \cdot \mathbf{B})/2\upi$ \citep{gj69}, one
can obtain analytically $f(\alpha) \approx i_{\rm s} + (i_{\rm a} - i_{\rm s})\sin^2\alpha$ with minimal assumptions
\citep{bes16, arz16}. We are grateful to Professor Vasily Beskin (Lebedev Physical Institute) for drawing our attention
to this point. }

The overall aim of our research is to derive the correction for the
classical estimator $B_{\rm md}$ on the basis of the state-of-the-art understanding of the properties and spin-down physics of isolated NSs.
We start from the spin-down torque (\ref{eq:torque_general}) assuming (\ref
{eq:spitkovsky_obliq}) as initially derived by \cite{spitkovsky06} and developed by \cite{tche13} and \cite{phil14}. Taking into account the existing
observational constraints on isolated NS masses and obliquity distributions and considering a representative subset of more than 20
realistic equations of state, we estimate the realistic uncertainties
that can be achieved in the measurement of a pulsar magnetic field.

We note that after this work was finished,  \cite{nikitina16} published the results of their re-calculation of $B$ for 376
pulsars using their original values of magnetic angles. However,
they used the classical `magneto-dipolar' spin-down law and did not
take into account either the scatter of masses over the NS population
or a realistic equation of state.

The paper is organized as follows. In Section~\ref{sect:method} we provide the
general equations for the magnetic field calculations and introduce
the refined formula. In Section~\ref{sect:parameters} the masses and obliquities of isolated NSs are discussed and a list of realistic equations of state is
described. These data are used in Section~\ref{sect:refine} to calculate the uncertainty of the proposed estimator. In Section~\ref{sect:discuss} the possible impact
of the pulsar timing irregularities on the magnetic field estimations,
as well as applications of our results, are discussed. In Section~\ref{sect:conclude}
a summary of the work is provided and conclusions are enumerated. Finally, in Appendix~\ref
{sect:eos_calc}, the parameters of a NS with various
equations of state are given and relevant calculations are presented.

\section{The timing-based magnetic field estimator}
\label{sect:method}
Hereafter, we assume that a radiopulsar with magnetic field strength
$B$ at the equator, spin period $P$ and obliquity $\alpha$ loses its rotational
energy according to the equation
\begin{equation}
    P\dot P = \dfrac{4\upi^2 R^6}{I c^3} B^2  (1 + 1.4\sin^2\alpha),
    \label{eq:spitkovsky}
\end{equation}
which was derived numerically for a spherical oblique rotator with
a plasma-filled magnetosphere by \cite{phil14} and is
nothing more than a restatement of the general equation (\ref{eq:spindown_general}) with $f(\alpha)$ in the form (\ref{eq:spitkovsky_obliq}) with 
$k_0 = 1.0$ and $k_1 = 1.4$. Extracting the value
of the magnetic field gives
\begin{equation}
    B = \sqrt{\dfrac{c^3}{4\upi^2}}  \dfrac{\sqrt{I(M)}}{R^3(M)} \sqrt{\dfrac{P \dot P}{1 + 1.4 \sin^2\alpha}}
    \label{eq:B_define}
\end{equation}
such that $B$ is dependent on the instantaneous angle $\alpha$, the equation
of state (EOS) of dense matter (i.e. the inertia and the size of the
star) and the full NS gravitational mass $M$

The difference between the logarithms of (\ref{eq:B_define}) and the classical `magneto-dipolar' estimator (\ref{eq:B0})
\begin{equation}
    \Delta_{\rm B} \equiv \log B - \log B_{\rm md}
\end{equation}
can be expressed as the sum of three terms:
\begin{equation}
    \Delta^{\rm (eos)}_{\rm B}(M, \alpha) = \Delta_{\rm eos}(M) + \Delta_{\rm
      oblq}(\alpha) + \Delta_{\rm norm} .
    \label{eq:DlogB}
\end{equation}
Here $\Delta_{\rm eos}(M)$ describes the deviation of the actual values of $I$ and $R$
from $I_0$ and $R_0$ respectively, namely:
\begin{equation}
    \Delta_{\rm eos}(M) = \dfrac12 \log \left[ \dfrac{I(M)}{I_0} \right] - 3\log\left[ \dfrac{R(M)}{R_0} \right],
    \label{eq:Deos}
\end{equation}
the second term is the obliquity correction
\begin{equation}
    \Delta_{\rm oblq}(\alpha) = -\dfrac{\log (1 + 1.4\sin^2\alpha)}{2},
    \label{eq:Dalpha}
\end{equation}
while the last term is the constant renormalization bias
\begin{equation}
    \Delta_{\rm norm} \equiv \dfrac{\log (2/3)}{2} \approx -0.088
\end{equation}
resulting from the difference in the numerical coefficients of the
classical and adopted versions of the spin-down laws. It can be seen
that the obliquity correction (\ref{eq:Dalpha}) has stringent boundaries owing
to $|\sin\alpha| \leq 1$. This term never exceeds zero for the adopted spindown model and reaches its minimal value $\approx$ -0.19 for a perfectly
orthogonal rotator ($\alpha = 90\degr$)

Thus, if the pulsar mass $M$, magnetic angle $\alpha$, spin period $P$ and its
derivative $\dot P$ are known, and if also the EOS is defined in the form of
$R(M)$ and $I(M)$ relationships, then the logarithm of the pulsar surface
field $B$ can be accurately calculated as $\log B_{\rm md} + \Delta_{\rm B}$, assuming that the spin-down rate follows Spitkovsky's law (\ref
{eq:spitkovsky}). However, $P$ and $\dot P$
are the only quantities from this list that can be routinely obtained
from observations.

The obliquity $\alpha$ is difficult to determine for an individual pulsar
\citep[e.g][]{nikitina11a}. However, these values are currently
known with low but still satisfactory accuracy for hundreds of
objects. Therefore, the distribution $p(\alpha)$ can be constructed with a
satisfactory precision (see Section \ref{sect:psr_obliq} for details).

The masses of {\it isolated} pulsars cannot be measured directly at
all. However, a few tens of mass estimations have been obtained
for slow, non-recycled pulsars in binary systems. These data are
believed to be relevant to isolated NSs as well, because slow pulsars
are expected to have masses near their birth values \citep[see the review by][and references therein, see also Section \ref{sect:psr_mass}]{ozel16}.
Thus, the distribution of the masses $p(M)$ of isolated NSs can also
be considered as known.

Note that the mass $M$ of an isolated pulsar is unlikely to be correlated with the instantaneous obliquity $\alpha$ or spin period $P$. Indeed, $M$
remains constant during the lifetime of a pulsar, while $\alpha$ and $P$ are
slowly evolving, decreasing and increasing respectively \citep{phil14}. Therefore, even if $M$ is strongly correlated with the birth
values $\alpha_0$ or $P_0$ (which, to our knowledge, has not been predicted
by any theoretical model), then the evolutionary de-correlation will
make these quantities statistically independent over the observed
pulsar population. Therefore, for our research we assume $\Delta_{\rm eos}(M)$ to
be statistically independent from both $\log B_{\rm md}(P, \dot
P)$ and $\Delta_{\rm oblq}(\alpha)$.

Moreover, we also assume that the correlation between $\Delta_{\rm oblq}(\alpha)$ and $\log B_{\rm md}(P, \dot P)$ (in other words, between $\alpha$ and $P$ or 
$\dot P$ ) is also weak and can be neglected. This statement is not so evident and
will be justified in detail in Section \ref{sect:psr_obliq} below.

The assumptions of the mutual statistical independence of all
components of $\Delta^{\rm (eos)}_{\rm B}$ and their consequent independence from
$\log B_{\rm md}$ allow the introduced correction to be described as a random quantity that follows a unified probability distribution function,
depending only on the adopted EOS:
\begin{equation}
    \Delta^{\rm (eos)}_{\rm B} \sim  p(\Delta_{\rm B}\vert{\rm eos})
\end{equation}
This distribution can be calculated numerically by substituting the
distributions of $\Delta_{\rm eos}(M)$ and $\Delta_{\rm oblq}(\alpha)$ (derived from $p(M)$ and $p(\alpha)$
respectively) and taking into account the $R(M)$ and $I(M)$ relationships (derived from the EOS). In this case, the basic moments of
$p(\Delta_{\rm B}\vert{\rm eos})$, the expectation
\begin{equation}
    \langle \Delta^{\rm (eos)}_{\rm B}\rangle = \int{\rm d}\alpha \int \Delta^{\rm (eos)}_{\rm B}(M, \alpha)  p(\Delta_{\rm B}
\vert{\rm eos}) {\rm d}M
\end{equation}
and variance
\begin{multline}
    \sigma^2[\Delta^{\rm (eos)}_{\rm B}] = \\ \int{\rm d}\alpha \int \left[\Delta^{\rm (eos)}_{\rm B}(M, \alpha) - \langle \Delta^{\rm
(eos)}_{\rm B}\rangle \right ]^2 p(\Delta_{\rm B}\vert{\rm eos}) {\rm d}M
	\label{eq:sigma_eos}
\end{multline}
have a clear physical meaning. In particular,
\begin{equation}
    \log B^{\rm (eos)}(P, \dot P) = \log B_{\rm md}(P, \dot P) + \langle \Delta^{\rm (eos)}_{\rm B}\rangle
    \label{eq:logBeos}
\end{equation}
provides an {\it unbiased}  estimation of the pulsar magnetic field with a
typical uncertainty of order $\sigma[\Delta^{\rm (eos)}_{\rm B}]$ for a given EOS. This equation
is the basic theoretical result of our research.

Its implementation may, however, be difficult, because the equation of state of NS matter remains formally unknown. Nevertheless,
a large number of reasonable theoretical propositions about it have
been made so far (see Section \ref{sect:eos} below). Adopting a number of
them, it is possible to construct a timing-based estimator of a pulsar magnetic field that is even more general than (\ref{eq:logBeos}). Indeed, let
$\{ {\rm eos}_i\}$ be a large enough subset of $N$ representative EOSs that do
not contradict the experimental data. Let also $w_i$ be the weight of the
$i$th EOS ($i = 1...N$) in the list estimating its chances to be realized
in nature, so that $\sum w_i
= 1$. Then a new random quantity $\Delta^*_{\rm B}$ can be
introduced through the mixture probability density

\begin{equation}
    p(\Delta^*_{\rm B}) = \sum_i w_i \times p(\Delta_{\rm B}\vert{\rm eos}_i)
\end{equation}
This quantity is nothing more than a general correction to the classical magnetic field value $\log B_{\rm md}$.
Its expectation
\begin{equation}
    \langle \Delta^*_{\rm B} \rangle = \sum_i w_i \times \langle \Delta^{\rm (eos_i)}_{\rm B}\rangle
\end{equation}
and variance
\begin{equation}
\sigma^2[\Delta^*_{\rm B}] = \sum_i w_i \times \left
(\sigma^2[\Delta^{\rm (eos_i)}_{\rm B}] + \langle \Delta^{\rm
(eos_i)}_{\rm B}\rangle^2 \right) - \langle \Delta^*_{\rm B}
\rangle^2,
\end{equation}
have the same meanings as the moments of $p(\Delta_{\rm B}\vert{\rm eos})$. Namely, the quantity
\begin{equation}
    \log B^*(P, \dot P) = \log B_{\rm md}(P, \dot P) + \langle \Delta^*_{\rm B}\rangle
\end{equation}
also provides an unbiased estimation of the surface magnetic field of a radiopulsar with an uncertainty of order
$\sigma[\Delta^*_{\rm B}]$ when no EOS can be absolutely preferred from the list of $N$ possibilities.

\section{Parameters of isolated neutron stars}
\label{sect:parameters}

\subsection{Mass distribution}
\label{sect:psr_mass}

To date, a few dozen well-constrained masses of NSs have been obtained from observations of binaries, in which the post-Keplerian orbital parameters of the system components can be extracted because
of highly precise timings\footnote{The actual lists of NS masses measurements can be found at URLs
\url{https://stellarcollapse.org/nsmasses} and \url{https://jantoniadis.wordpress.com/research/ns-masses/}} \citep[e.g][]{lat12,
ant_thesis, ozel16}. Some of these data can represent the mass
distribution of isolated radiopulsars.

There are three types of compact binaries wherein the accretion episode was weak and/or relatively short term. These are double neutron stars (DNSs), binaries containing slow-rotating pulsars
with spin periods up to hundreds of seconds, and high-mass X-ray
binaries (HMXBs) with non-recycled pulsars. Following \cite{ozel16} -- OF16 hereafter -- we will consider the two latter
types together and refer to them as `slow pulsars'. Owing to the
weak impact of mass transfer, the NSs in these systems have relatively longer spin periods and stronger magnetic fields. They are
also expected to have systematically lower masses, namely close to
those at birth. Thus these systems are expected to be relevant for
the reconstruction of the mass distribution of isolated objects.

The accurate Bayesian analysis of the NS mass distribution has a
long history \citep[e.g][]{finn94, thor99, swa10}. The most recent results in this
topic have been obtained by \cite{ozel12},
\cite{kiz13} and OF16. It has been found that the NS mass probability density $p(M)$ can be well described by a Gaussian with the mean and
variance depending on a specific type of binary.

\begin{table}
\centering
\caption{The parameters of the neutron star mass distributions obtained by
various authors in the form of $\langle M/M_{\sun} \rangle \pm \sigma[M/M_{\sun}]$. DNS refers to the components
of double neutron stars. The numbers of pulsars involved in the analysis
(second column) are given as $N_{\rm _{DNS}}
+ N_{\rm slow}$. The distribution adopted within
this work is in boldface.}
\label{tab:nsmass}
\begin{centering}
\begin{tabular}{lccc}
  \hline
  Work & Pulsars & DNS & ``slow'' pulsars \\
  \hline
	\cite{ozel12} & 12 + 8 & $1.33 \pm 0.05$ & $1.28 \pm 0.24$\\
	\cite{kiz13} & 18 + 0& $1.33 \pm 0.12$ & n/a \\
	\cite{ozel16} & 22 + 12 & $1.33 \pm 0.09$ & $\mathbf{1.49 \pm 0.19}$ \\
	\hline
\end{tabular}
\end{centering}
	\label{tab:masses}
\end{table}

The parameters of the NS mass distributions obtained in these
papers are collected in Table~\ref{tab:masses}. It can clearly be seen that the masses
of DNS components have a significantly narrower distribution but are still well within the mass range of slow pulsars. This divergence
is not well understood as yet, but is likely to be the result of a
specific `fine-tuning' in the evolution of massive binary systems --
the progenitors of DNSs \citep
[e.g.][OF16]{postnov14}. So
the birth masses of NSs are nevertheless expected to be distributed
quite widely.

On the other hand, the masses of the {\it recycled pulsars} (known
from studies of NS-white dwarf systems) are consistent with a
normal distribution with $\langle M \rangle \approx 1.5-1.55M_{\sun}$ and $\sigma\sim
0.2M_{\sun}$ \citep[][OF16]{kiz13}. In other words, the recycled pulsars
that experienced a significant mass and momentum transfer appear
to be $\sim 0.2-0.25M_{\sun}$ more massive than components of DNSs,
but only barely exceed the masses of slow pulsars. This fact is in
tension with the prediction about the amount of mass accreted by a
NS within a low-mass X-ray binary (LMXB): $\Delta
M_{\rm acc} \lesssim 0.1-0.2 M_{\sun}$ \citep[e.g.][]{kiz13}. This estimation, however, was obtained
adopting quite a long accretion period of $\sim$10 Gyr. Moreover, the
recent analysis by \cite{anto16} has provided evidence
for a possible bimodality of $p(M)$ for recycled pulsars, with the
components consistent with Gaussians located at $1.39\pm 0.06$ and $1.81\pm 0.15M_{\sun}$ respectively (the numbers after `$\pm$' denote the
widths of the components). The existence of recycled pulsars whose
masses are concentrated near $\sim 1.8M_{\sun}$ is probably an indication
of quite a wide distribution of initial NS masses with a mean of
$\approx 1.5-1.6M_{\sun}$.

Therefore, we finally consider the distribution $p(M)$ obtained by
OF16 for slow pulsars to be the most conservative approximation
of the masses of isolated NSs:

\begin{equation}
    p(M/M_{\sun}) \propto \exp\left\{ - \dfrac{1}{2}\left(\dfrac{M/M_{\sun} -
          1.49}{0.19}\right)^2 \right\} ,
    \label{eq:mass_distrib}
\end{equation}
and adopt it in the calculations below. This distribution is quite wide,
permitting NSs with both `low' ($\sim 1.2M_{\sun}$) and `high' ($\sim 1.8M_{\sun}$)
masses. (It is shown in Fig.~\ref{fig:eos_list}(a) by the thick line with its $3\sigma$
boundaries, along with the distribution of the components of DNSs
derived in the same work.)

\subsection{Pulsar obliquities}
\label{sect:psr_obliq}

\begin{figure}
    {\centering \resizebox*{1\columnwidth}{!}{\includegraphics[angle=0]
    {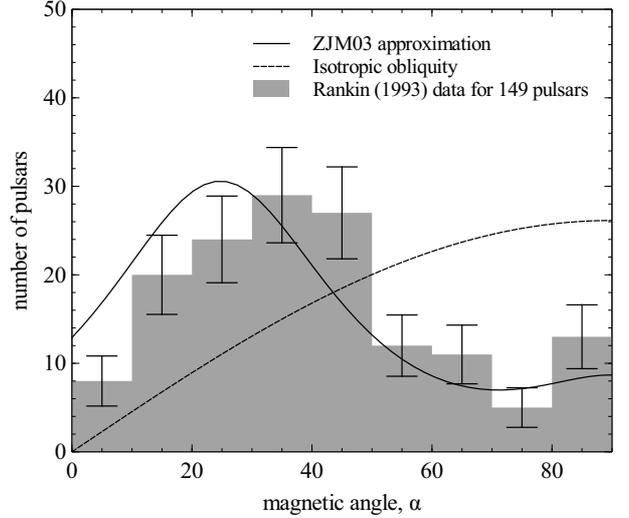}} \par}
    \caption{The distribution of magnetic angles for 149 normal pulsars published by\citet{rankin93a} (grey bars) and the statistical approximation (\ref	
	{eq:alpha_distrib}) found by ZJM03 (solid line). The error bars of the empirical distribution are
	from Poissonian statistics. The isotropic obliquity distribution (\ref{eq:alpha_iso}) is also
	shown (dashed line).}
    \label{fig:alpha_distrib}
\end{figure}

The angle $\alpha$ between the NS magnetic and spin axes has been determined for hundreds of isolated pulsars. The most confident values
have been obtained for 300+ pulsars by \cite{lyne88}, \cite{rankin93a, rankin93b} and \cite{gould94}, using various methods. \cite{tm98} combined all the measurements reported in
these papers and provided an extensive analysis of their statistics.
The estimations made by \cite{rankin93a} were found to be fully
consistent with the results published by other authors. Their distribution is shown in Fig.~\ref{fig:alpha_distrib}. As can clearly be seen, the apparent $p(\alpha)$ is
unlikely to be isotropic, but there is evidence for a secular magnetic
alignment so that most of the pulsars tend to have $\alpha \lesssim 45\degr$.

By performing an accurate statistical analysis of this data set,
\cite{zjm03} (ZJM03 hereafter) derived an analytic equation describing the apparent $p(\alpha)$:
\begin{equation}
    p(\alpha) = \dfrac{0.6A}{\cosh(3.5(\alpha - 0.43))} + \dfrac{0.15A}{\cosh(4(\alpha - 1.6))}
    \label{eq:alpha_distrib}
\end{equation}
where $\alpha = 0-\upi/2$ rad, and $A \approx 1.96$ is the
normalization constant. We have adopted this model for the calculations below. At the same
time, we also checked out the more conservative assumption about
the isotropy of pulsar obliquities such that
\begin{equation}
    p(\alpha) = \sin\alpha.
    \label{eq:alpha_iso}
\end{equation}
The important point, however, is that Spitkovsky's spin-down
model predicts the evolutionary decrease of $\alpha$ on time-scales of
a few $\times$ $10^7$ years. Moreover, this process has been indirectly
confirmed in observations with a number of statistical methods
\citep[e.g][]{xuwu91, tm98, welt08, young10}. Therefore one can naturally expect a non-zero correlation between the obliquity correction $\Delta_{\rm oblq}$
and the `magnetodipolar' estimator $B_{\rm md} \propto \sqrt{P \dot P}$ over the subset of
observed radiopulsars.

\begin{figure}
    {\centering \resizebox*{1\columnwidth}{!}{\includegraphics[angle=0]
    {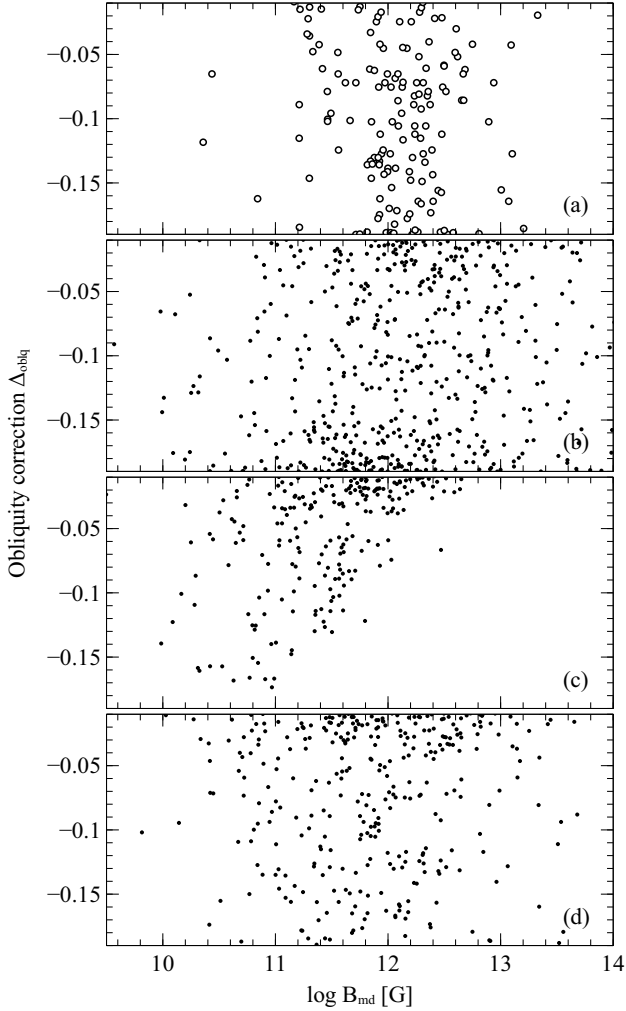}} \par}
    \caption{The dependence of the obliquity correction$\Delta_{\rm oblq}$ on $\log B_{\rm md}$.
(a) The apparent correlation for 149 pulsars with obliquities measured by
\citep{rankin93a}. (b) The simulated values $\Delta_{\rm oblq}$ (\ref{eq:Dalpha}) and $\log B_{\rm md}$ in the
case of isotropic obliquities $\alpha_0$ at pulsar birth (see text for details of the
simulations). (c) The same, but for a perfect positive correlation between $\alpha_0$
and the initial periods $P_0$. (d) The same, but for a perfect negative correlation
between $\alpha_0$ and $P_0$.}
    \label{fig:obliq_correlation}
\end{figure}

However, \cite{tm98} found that the values of
$P$ and $\alpha$ for 300+ pulsars do not show any significant correlation,
which is probably the result of a large scatter of $\alpha$ for a given period $P$. In our work, we have also investigated the direct correlation
between $\Delta_{\rm oblq}$ and $\log B_{\rm md}$ for 149 pulsars with known obliquities
\citep{rankin93a} by plotting this correlation in Fig.~\ref
{fig:obliq_correlation}(a). The independence of these parameters is clearly seen there. It can be simply
understood by assuming a weak correlation between the magnetic
obliquity $\alpha_0$ and period $P_0$ at the pulsar birth and taking into account
the weak dependence of the instantaneous $\dot P$ on $\alpha$, as predicted by
the spin-down law (\ref{eq:spitkovsky}).

In addition, we have undertaken numerical simulations of $\Delta_{\rm oblq}$ and $\log B_{\rm md}$ by solving the equations of pulsar spin-down using the full braking torque proposed by \cite{phil14}. In particular,
the obliquity evolution equation
\begin{equation}
	I\Omega\dfrac{d\alpha}{dt} = - \dfrac{\mu^2\Omega^3}{c^3}\sin\alpha\cos\alpha
	\label{eq:alpha_evolution}
\end{equation}
has been solved simultaneously with (\ref{eq:spitkovsky}) for $10^3$ synthetic pulsars.
Initial periods $P_0$ were from a normal distribution centred on 300 ms
with standard deviation 150 ms \citep{fgk06},
magnetic moments $\log \mu$
[G cm$^3$] $\sim {\rm normal}(30.5,
1.0)$, while ages
$t $were from a uniform distribution over the interval [0; 100Myr]. At
the same time, while NS radii were set to 12 km for all simulated
pulsars, their masses $M$ were calculated as random values according
to the distribution (\ref{eq:mass_distrib}) with the most probable values for the moment
of inertia $I = 10^{45} \cdot (M/M_{\sun})$ g cm$^2$ (see Fig.~\ref{fig:eos_list}c).

The results of the simulations are shown in Fig.~\ref{fig:obliq_correlation}(b)-(d). These
three plots correspond to the different initial conditions adopted for
the modelling: (2b) initial obliquities $\alpha_0$ are distributed isotropically
and independently from initial periods $P_0$; (2c) $\alpha_0$ are perfectly
correlated with $P_0$ so $\sin\alpha_0 = {\rm cdf}(P_0/{\rm sec})$; (2d) $\alpha_0$ are perfectly anti-correlated with $P_0$ so $\sin\alpha_0 = 1 - {\rm cdf}(P_0/{\rm sec})$. Here ${\rm cdf}(x)$ is
the cumulative distribution function of an initial period:

\begin{equation}
	{\rm cdf}(x) = \dfrac{1}{2}\left[ 1 + {\rm erf} \left( \dfrac{x - 0.3}{0.15\sqrt{2}}\right) \right ]
\end{equation}
and ${\rm erf}(\cdot)$ is the error function.

It has been obtained once again that $\Delta_{\rm oblq}$ do not correlate
with $\log B_{\rm md}$ in all cases, except for the model in which a strong positive dependence $\alpha_0(P_0)$ was adopted (Fig.~\ref{fig:obliq_correlation}c). Even in this
case, however, the apparent correlation is very weak. Therefore,
$\Delta_{\rm oblq}$ and $\log B_{\rm md}$ can be considered to be statistically independent
for actual pulsars. By combining this result with the statement about
the independence of $\Delta_{\rm eos}$
 and $\log B_{\rm md}$ one finds the accurately calculated correction $\Delta_{\rm B}$ to be uncorrelated with the classical estimator
$\log B_{\rm md}$. Hence, the width of the distribution of $\Delta_{\rm B}$ can be considered
to be the same as that of the corrected value $\log B_{\rm md} + \Delta_{\rm B}$, making
$\sigma[\Delta_{\rm B}]$ (\ref{eq:sigma_eos}) the statistical uncertainty of the refined estimation of the
magnetic field.

\subsection{Realistic equations of state}
\label{sect:eos}

The final step is the compilation of a representative subset of realistic equations of state. The recent mass measurements for pulsars
 PSR J1614-2230  ($1.97\pm 0.04M_{\sun}$,
\cite{Demorest}) and  J0348+0432 ($2.01\pm 0.04 M_{\sun}$
\cite{Antoniadis}) ruled out
many of the EOSs proposed earlier. In particular, for various EOSs
including hyperons, the maximal mass limit for non-magnetic NSs
is considerably lower than two solar masses.

There are some indications that two binaries -- B1957+20 \citep{Kerk} and
4U 1700-37 \citep{clark02} -- contain even more massive NSs with
masses $\sim 2.4M_{\sun}$, even though the systematic errors of mass measurements are large. Taking into account the aforementioned examples, we excluded from 
our consideration EOSs for which the
maximal mass of a NS is below $1.95M_{\sun}$ as being unrealistic.

\begin{figure}
    {\centering \resizebox*{0.99\columnwidth}{!}{\includegraphics[angle=0]
    {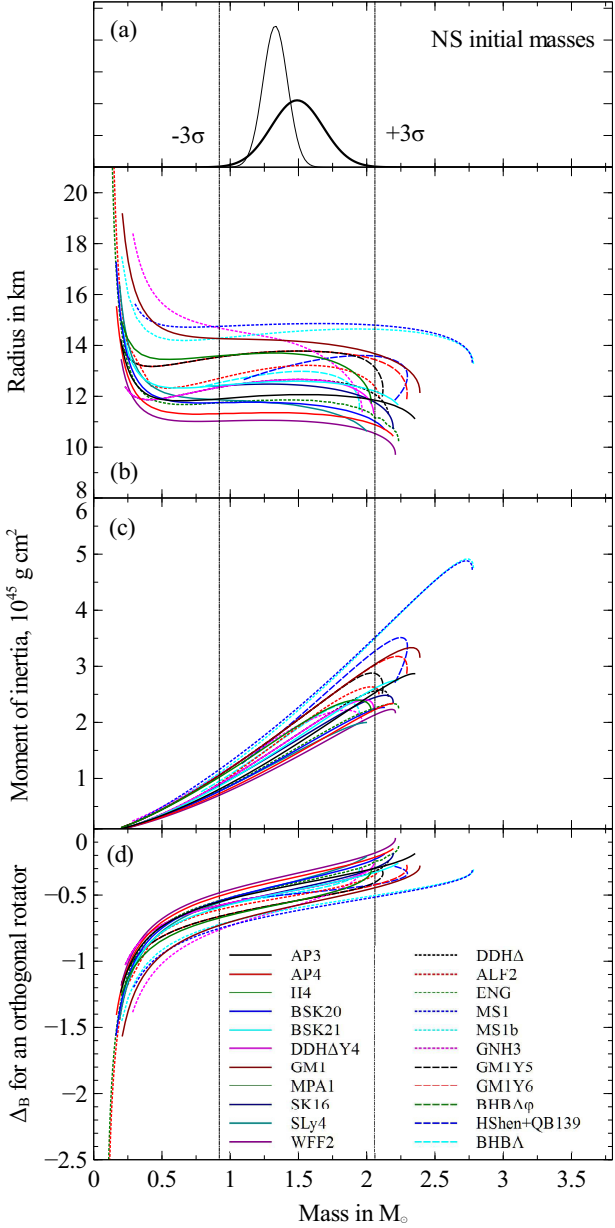}} \par}
    \caption{The subset of 22 realistic equations of state of dense matter
selected for the analysis. (a) The distribution of the initial masses of neutron
stars (equation~\ref{eq:mass_distrib}) with its $\pm 3\sigma$ boundaries is shown by the thick line, while
the thin line represents the corresponding distribution for the components
of double neutron stars as derived by OF16. (b) The classical mass--radius
diagram. (c) Neutron star moment of inertia versus the full mass relation.
(d) The correction $\Delta_{\rm B}$ (equation~\ref{eq:DlogB}) calculated for an orthogonal rotator, that
is, for constant $\Delta_{\rm oblq} \approx
-0.19$. Its value is always negative for all the EOSs
and lies within the interval -0.75 to -0.25 for most probable values of
masses according to (\ref{eq:mass_distrib})}
    \label{fig:eos_list}
\end{figure}

Unfortunately, there are no high-accuracy measurements of NS
radii, and, moreover, there are no such measurements at all for any
NS with a precise mass determination. Fortunately, some knowledge of the mass--radius relationship can be derived from observations and the modelling of X-ray bursts. In particular, from observations of XTE
J1807-294, SAX J1808-3658, and XTE J1814-334 it can be concluded that in the wide interval of masses
$1M_{\sun} < M < 2.3M_{\sun}$ the radius for NSs should be approximately
constant, $\sim 12 $ km \citep{Leahy}. Other investigations \citep{SLM, HAMB} of longer
X-ray bursts, however, witness in favour of larger radii $R\geq 14$ km.
We assumed the values $11.0<
R_{1.4}<15$ km for the radius of a
$1.4M_{\sun}$ NS and therefore excluded from our calculations EOSs
with more compact or larger stars.

For completeness of our analysis, it is necessary to consider
various classes of EOSs as distinguished by the approaches used
for their construction. Most EOSs fall into one of three groups, as
follows.

(1) EOSs obtained from non-relativistic many-body calculations.
The well-known SLy4 EOS (\citep{SLy, SLy-4} uses a simple model of two-nucleon potential and is
based on a single effective nuclear Hamiltonian. We also considered
the WFF2 EOS based on the Urbana V14 two-nucleon potential
\citep{WFF} with a maximal mass of star
exceeding 2$M_\odot$.

Models with three-nucleon interactions are based on data for energies of light nuclei and/or properties of symmetric nuclear matter (SNM) with a comparable number of neutrons and protons.
We considered AP3 and AP4 EOSs  \citep{APR}. These equations were obtained using variational
techniques and the Argonne 18 potential plus a three-body UIX
potential (AP3) and A18+UIX potentials with $\delta v$ relativistic boost
corrections (AP4).

The next two unified EOSs for cold catalysed nuclear matter developed by the Brussels-Montreal group (BSK20 and BSK21, see
\cite{BSKEOS}; \cite{BSKEOS-1} and \cite{BSKEOS-2}) are calculated using the TETFSI method
(temperature-dependent extended Thomas-Fermi plus Strutinsky
integral) for functionals based on Skyrme-type forces. Recently,
\cite{SK16} proposed a unified EOS (SK16) using
the effective interaction model and cluster energy functionals from
\cite{SK16-1, SK16-2}.

(2) EOSs calculated from the relativistic Dirac-Brueckner-
Hartree-Fock (DBHF) approach to dense neutron matter. We considered two EOSs of this type, namely ENG \citep{ENG}
and MPA1 \citep{MPA} with the inclusion
of contributions from the exchange of $\pi$ and $\rho$ mesons.

\begin{figure*}
    {\centering
    \resizebox*{1\columnwidth}{!}{\includegraphics[angle=0] {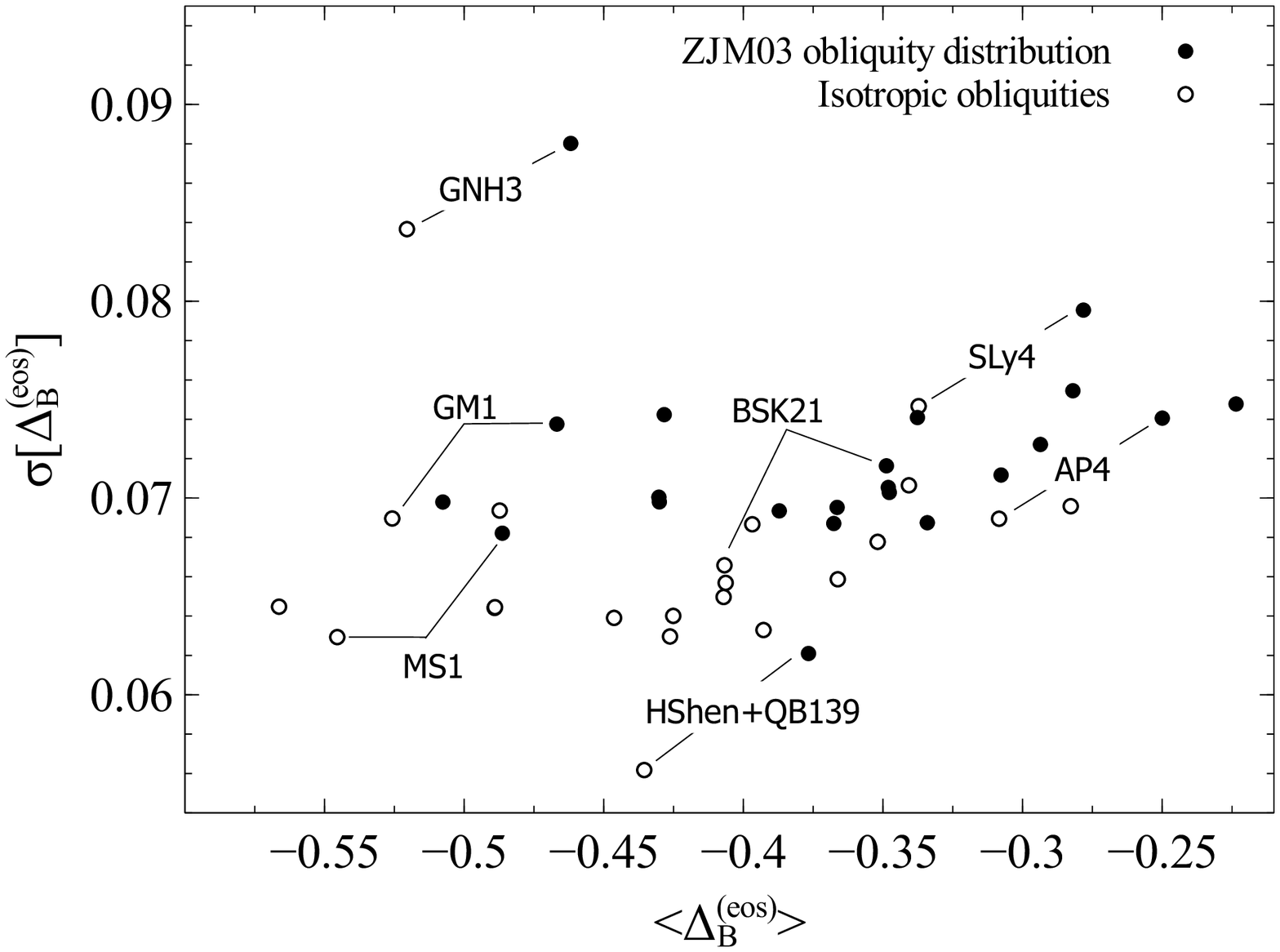}}
    \resizebox*{1\columnwidth}{!}{\includegraphics[angle=0] {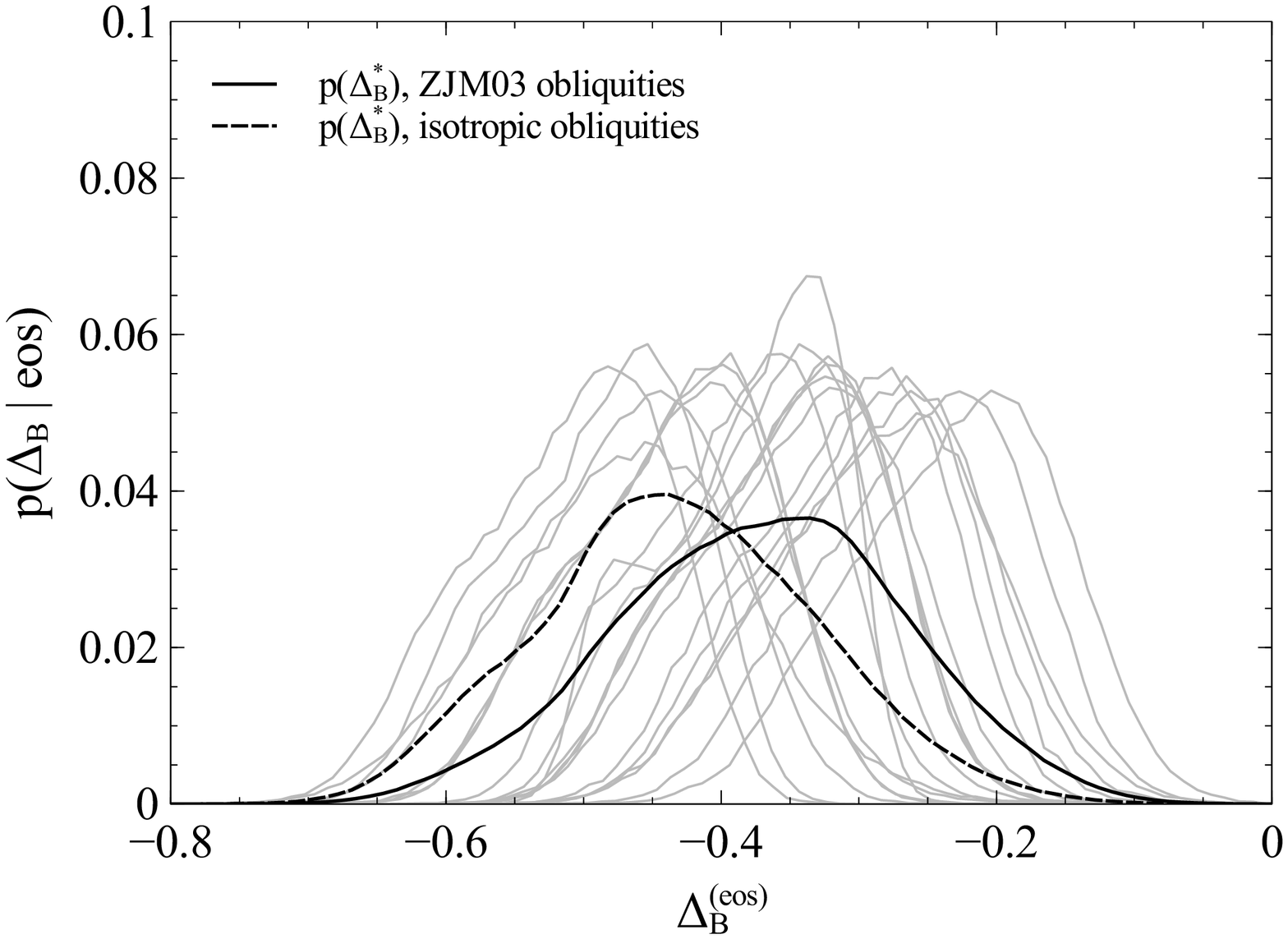}}
    \par}
    \caption{(Left) The basic moments -- expectation $\langle \Delta^{\rm (eos)}_{\rm B} \rangle$ and standard deviation $\sigma[\Delta^{\rm (eos)}_{\rm
B}]$ -- of the distributions of correction $\Delta^{\rm (eos)}_{\rm B}$ for the subset of 22 equations of state (EOSs). The values represented by solid circles 
were calculated adopting the distributions of magnetic obliquity and neutron star masses described by (\ref{eq:alpha_distrib}) and (\ref{eq:mass_distrib}) 
respectively. Open circles, in turn, assume the model (\ref{eq:alpha_iso}) of isotropically distributed obliquities. Some points are labelled with the
name of the corresponding EOS. It is clear that, while the average of $\Delta_{\rm B}$ covers the interval from $\approx -$0.55 to $\approx -$0.25, its dispersion 
remains approximately the same for all EOSs with the typical value $\sigma[\Delta^{\rm(eos)}_{\rm B}] \sim 0.07$. This value represents the reachable precision 
of a timing-based estimation of a pulsar magnetic field within unknown obliquity and the mass of a neutron star. (Right) The distributions of 
$p(\Delta_{\rm B}\vert{\rm eos})$ for 22 equations of state within the ZJM03 obliquities (grey lines). The distributions of the generalized correction 
$p(\Delta^*_{\rm B})$ adopting the ZJM03 model and isotropic obliquities are shown by the solid and dashed black curves respectively. All the plotted 
distributions are roughly close to the normal one.}
    \label{fig:delta_distribs}
\end{figure*}

(3) EOSs from relativistic mean-field theoretical models (RMFs).
In particular, we take for calculations the well-known GM1 EOS
\citep{GM}). This EOS is based on the classical parametrization proposed by Glendenning and Moszkowski.
More realistic models predict the existence of hyperons in NS cores
at densities of $5-8\times 10^{14}$ g/cm$^{3}$. Two extensions of the GM1
model, GM1Y5 and GM1Y6 \citep{GM-1}, are obtained for
cold NS matter in $\beta$-equilibrium containing the baryon octet and
electrons.

Another RMF parametrization, DDH$\delta$, was considered by
\cite{DDH}. Its extension on the hyperon (DDH$\Delta$4 EOS)
sector in the same manner as for the GM1 model is proposed in
\cite{DDH-2}. BHB$\Lambda$ EOS and its variation  BHB$\Delta\phi$ EOS
\citep{BHB} are obtained from the
statistical model of \cite{BHB-2} with RMF
parametrization DD2 \citep{BHB-3}, extended by $\Lambda$-hyperons
or $\Lambda$-hyperons interacting via the $\phi$-meson. We also consider EOSs
proposed by Mueller and Serot \citep{MS} (MS1, MS1b)
taking account of the non-linear interactions between the scalar-isoscalar ($\sigma$), vector-isoscalar ($\omega$), and vector-isovector ($\rho$) fields

The family of GNH EOSs with nucleon-hyperon composition
was proposed by \citep{GNH}. We use the GNH3 EOS with
universal couplings for hyperons in our calculations. This EOS gives
an acceptable upper limit of mass for a star.

Seven hyperonic EOSs in relativistic mean field theory were proposed by \cite{H4}. Only one of these equations --
namely H4 -- is consistent with the two-solar-mass limit. In contrast to the GNH3 EOS, the hyperon-meson couplings for H4 are
assumed to be the same for all hyperons but are weaker than the
nucleon-meson couplings. The stiffness is achieved owing to the
high value of incompressibility, K = 300 MeV

It cannot be ruled out that some compact stars are `hybrid stars'
with cores consisting of quark matter. However, many such EOSs
are not compatible with the upper mass limit and radius constraints.
Two hybrid EOSs with nuclear matter and colour-flavour-locked
quark matter are included in our consideration. The ALF2 EOS
was proposed by \cite{ALF2}. As shown for acceptable
parameters of the MIT bag model with gluonic corrections, one can
obtain a mass--radius relationship for hybrid stars similar to that
predicted for stars from nucleonic matter. For nuclear matter, the
EOS proposed by Akmal, Pandharipande and Ravenhall (APR) was
used

The HShen+QB139 EOS \citep{HSHEN, HSHEN-2, HSHEN-3} is based on the model proposed by \cite{HSHEN-4}
with the addition of a bag model for the quark phase. The transition
to quark matter has been described in the frame of the non-linear
RMF model with TM1 parametrization \citep{HSHEN-5}.

Thus, we have finally compiled a list of 22 EOSs that we consider
to be representative enough for the current state of knowledge. The
basic parameters of NSs as well as the mass--radius and mass--
moment of inertia relationships were calculated for each equation.
The basics of EOS calculus can be found in Appendix ~\ref
{sect:eos_calc}, while
the results are shown in Table~\ref{tab:NSC} and in Fig.~\ref{fig:eos_list}(b) and (c). The
correction $\Delta_{\rm B}$ in the case of an orthogonal rotator ($\alpha = 90\deg$) was
also calculated and is shown in Fig.~\ref{fig:eos_list}(d).

It can be seen that EOSs from the DBHF approach give results
similar to non-relativistic EOSs from many-body calculations (these
EOSs are listed in Table~\ref{tab:NSC} above the horizontal line). The radii for
a NS with canonical mass $1.4M_\odot$ vary from 11.04 km (WFF2) to
12.58 km (BSK21), and the moment of inertia of such a star lies in
a narrow interval $1.28 < I(1.4M_\odot) < 1.57\times
10^{45}$ g cm$^{2}$. For RMF
EOSs we have relatively larger radii, $R\sim 13-14$ km, and the interval for the moment of inertia $1.74 < I(1.4M_\odot) < 2.05\times 10^{45}$ g cm$^{2}$.
We also include in the table the results for stars with $M=1.49M_\odot$.
Therefore, the canonical value for $I(1.4M_\odot)=10^{45}$ g cm$^{2}$ seems
to underestimate the actual values of NSs moments of inertia for
30-60 per cent at least.

\section{Properties of the refined estimator}
\label{sect:refine}

We have calculated numerically the distributions of $\Delta^{\rm
(eos)}_{\rm B}$ and their
basic moments for all the EOSs described above. The values of $M$
and $\alpha$ have been simulated $10^5$ times for each run according to the
empirical distributions of these quantities discussed in Sections ~\ref{sect:psr_mass}
and~\ref{sect:psr_obliq} respectively.

The results are shown in Fig.~\ref{fig:delta_distribs}. Adopting the ZJM03 model of
pulsar obliquities, the shape of $p(\Delta_{\rm B}\vert{\rm eos})$ was found to be very close
to the Gaussian for every EOS (see the right panel of the figure).
The corresponding means$\langle \Delta^{\rm(eos)}_{\rm B } \rangle$ cover a relatively narrow interval
of values from $\approx -0.51$ (for MS1) to $\approx -0.23$ (for WFF2), while
the standard deviations $\sigma[\Delta^{\rm(eos)}_{\rm B}]$ appear to be nearly the same for
all equations:

\begin{equation}
    \sigma[\Delta^{\rm (eos)}_{\rm B}] \approx 0.07\pm 0.01 .
    \label{eq:sigma_delta_b}
\end{equation}
The latter result seems to be the consequence of two factors. The
first is the existence of a flat plateau in $R(M)$ relationships, which
is typical for the considered EOSs. In other words, all isolated NSs
are likely to have the same radii within the adopted distribution
of their masses. The second factor is that the pulsar spin-down
luminosity weakly depends on the instant obliquity $\alpha$, which makes
the distribution of $\Delta_{\rm oblq}(\alpha)$  quite narrow. Indeed, when adopting the
spin-down law that is strongly dependent on $\alpha$ ($x_{\perp} = 2/3$, $x_{\parallel} = 0$ so $P\dot P \propto B^2 \sin^2\alpha$ in the terms of equations (\ref{eq:f_general}) and (\ref{eq:spindown_general})), we obtained
typical $\sigma[\Delta^{\rm(eos)}_{\rm B}] \approx 0.36$, which is five times greater than (\ref{eq:sigma_delta_b}).

On the other hand, if the isotropic distribution of $\alpha$ is used, the
distributions $p(\Delta_{\rm B}\vert{\rm eos})$ keep their shapes and widths in general. The
corresponding means $\langle \Delta^{\rm(eos)}_{\rm B } \rangle$ are just shifted by $\sim$0.05 to the left
relative to the points given by the ZJM03 model (as shown by open
circles in the same panel). So, we conclude that the specific choice
of $p(\alpha)$ does not affect significantly the results of the calculations
of the correction $\Delta_{\rm B}^{\rm (eos)}$.

Finally, the distribution of the generalized correction $\Delta^*_{\rm B}$ under
the assumption of equal weights (we used $w_i = 1/22$) of every EOS
from our list was also calculated. The resulting curves are shown
with the black solid (for ZJM03) and dashed (for isotropic $p(\alpha)$)
lines in the right panel of Fig. 4. For the ZJM03 model of pulsar
obliquities we found that

\begin{equation}
    \log B^{\rm *} - \log B_{\rm md} \approx -0.37 \pm 0.10 ,
\end{equation}
while isotropic model gave
\begin{equation}
    \log B^{\rm *} - \log B_{\rm md} \approx -0.43 \pm 0.10 .
\end{equation}
This result means that the timing-based estimation of a pulsar magnetic field can be as precise as  $\sim 0.1$ dex even if neither the EOS
nor the mass nor the obliquity is known. It is an intrinsic property
of the adopted pulsar spin-down luminosity model. If it is rewritten
in a linear form, the quantity
\begin{equation}
    B^* \approx \frac37 B_{\rm md}
\end{equation}
provides an unbiased estimation of the actual surface magnetic
field strength of an isolated radiopulsar with only $\sim$20-25 per cent
uncertainty at the 68 per cent confidence level.

\section{Discussion}
\label{sect:discuss}

\subsection{Do pulsars timing irregularities affect $\log B$?}
\label{sect:timingnoise}
We have hitherto assumed that the instantaneous NS magnetic field
$B$ is linked to the {\it observed} $P$ and $\dot P$ through the spin-down law
(\ref{eq:spitkovsky}) in the strict sense. However, the actually observed spin evolution of isolated pulsars is more complex than predicted by that
equation. The spin-down rate appears to be significantly contaminated by additional irregular, quasi-periodic variations on short
time-scales from months to years that are typically referred to as
the `timing noise' \citep{boyn72, ch80, daless95, urama06, hobbs10, nice13}.

The strength of this unmodelled component of the pulsar spindown can be characterized numerically with the dimensionless combination
\begin{equation}
	n = \dfrac{\ddot \Omega \Omega}{\dot \Omega^2}
	\label{eq:bi_obs}
\end{equation}
-- the so-called ``braking index'. Differentiating the general spin-down law (\ref{eq:spindown_general}), gives
\begin{equation}
	n = 3 + \dfrac{\Omega}{\dot \Omega} \left [2\dfrac{1}{\mu}\dfrac{{\rm d}\mu}{{\rm
d}t} - \dfrac{1}{I}\dfrac{{\rm d}I}{{\rm d}t} + \dfrac{1}{f}\dfrac{{\rm d}f}{{\rm d}\alpha} \dot\alpha \right],
	\label{eq:bi}
\end{equation}
where the first term reflects the power of the $\Omega$ term in the braking torque equation (\ref{eq:torque_general}). In other words,
if the spin-down of a NS follows the expression
$\dot \Omega = - K \Omega^q$, then the combination
(\ref{eq:bi_obs}) is simply equal to $q$ assuming constant $K$.

Thus, $n \in 3..3.25$ for the Spitkovsky's law when magnetic moment $\mu$ and moment of inertia $I$ are fixed but
the obliquity $\alpha$ decreases according to (\ref{eq:alpha_evolution}) \citep[e.g.][]{eksi16}. Other models for $f(\alpha)$ and
$\dot\alpha$ are able to significantly extend this interval up to $0 \la n \la 10^4$. For instance, the classical magneto-dipolar
losses ($x_{\perp} = 2/3$, $x_{\parallel} = 0$) give $n = 3 + 2\cot^2\alpha \la 10^3$ for actual pulsars.

Nevertheless, the estimated values of $n$ for hundreds of pulsars are
surprisingly far from any prediction. Their values have been found in
the range from $\sim -10^6$ to $\sim 10^6$, being negative for about half of the
objects \citep[e.g][]{hobbs04, bbk12, zhang12}. So, they are unlikely to represent the secular
(i.e. evolutionary) spin-down of radiopulsars.

Note that the braking indices of most of the sources are not
stable from observation to observation. Even for a few tens of `lownoise' pulsars, however, the values of $n$ are constant within spans of
10-30 years but still extremely anomalous \citep{bbk07}.
Thus, finally, only 10 sources are accepted to date as having meaningful and stable $n \in 0.03$-$3.15$ that can be interpreted in terms of
the spin-down law (\ref{eq:spindown_general}) \citep{archi16, marsh16}.

Although the physics of the pulsar timing irregularities remains
generally unclear, the proposed solutions of the `anomalous braking
indices' problem can be qualitatively divided into two categories. Type I models incorporate the relatively slow variability of NS
parameters directly in the spin-down equation, assuming the variability of the surface magnetic field \citep
{pons12, zhang12, ou16}, obliquity \citep{melatos2000, lyne13, arz15} or effective moment of inertia \citep{tsang13, hamil15,
hamil16}. Such models strictly keep
the relationship between the observed $P$, $\dot P$ and $B$ in the form of the
adopted spin-down law at any moment of time. Hence, the timing-based magnetic field estimation $B \propto \sqrt{P \dot P}$ cannot be affected by
the processes of such a type.

Type II models suggest the existence of an {\it additional} either
quasi-periodic or purely stochastic component $\delta \dot \Omega$in the spindown rate. The underlying physics was proposed to be probably
connected with the magnetospheric perturbations \citep{cheng87, kramer06, cont07, lyne10}, or processes in the star interior \citep{janssen06,
ml14}, or to be the imprint of the so-called `anomalous braking
torque' \citep{bbk07, bars10, bbk12}. Within this approach, the observed spin frequency
derivative $\dot \Omega_{\rm obs}$ is the sum of $\dot \Omega \propto B^2\Omega^3(1 + 1.4\sin^2\alpha)$ and the
variational term

\begin{equation}
	\dot \Omega_{\rm obs} = \dot \Omega + \delta \dot \Omega = \dot \Omega \left (1 + \varepsilon \right),
\end{equation}
where $\varepsilon(t) = \delta\dot \Omega/\dot \Omega$ is the relative divergence. Note, that the corresponding divergence of
the spin frequency $\delta \Omega = \Omega_{\rm obs} - \Omega$ is assumed to be vanishingly small and can be neglected.

Generally, the wide range of the models proposed for $\delta\dot\Omega$ cannot
explain the properties of the observed braking indices completely
 \citep{malov16}. Formally, Type II solutions require the correction
$\Delta^{\rm (eos)}_{\rm B}$ (\ref{eq:DlogB}) to be extended by the additional term
\begin{equation}
	\Delta_{\varepsilon} = -\dfrac{\log(1 - \varepsilon)}{2}
	\label{eq:delta_eps}
\end{equation}
because $\delta \dot P/\dot P = -\delta \dot \Omega/\dot \Omega = -\varepsilon$. 
Moreover, because it is not possible to reject the hypothesis that both types of physical processes
(I and II) can contribute to the unmodelled part of the observed
spin-down, the term $\Delta_{\varepsilon}$ has to be added to $\Delta^{\rm (eos)}_{\rm B}$ as
\begin{equation}
	\Delta^{\rm (eos)}_{\rm B,ext} = \Delta^{\rm (eos)}_{\rm B} + r_{\rm ext}\Delta_{\varepsilon},
	\label{eq:logBext}
\end{equation}
where $r_{\rm ext} \in 0...1$ characterizes the `fraction' of the unmodeled spin-down from the Type II (i.e. external)
processes.  For instance, $r_{\rm ext} = 0$ means that anomalous braking indices can be explained by the $\mu$, $I$ and/or $\alpha$
variations only using equation (\ref{eq:bi}).

\begin{figure}
    {\centering \resizebox*{1\columnwidth}{!}{\includegraphics[angle=0]
    {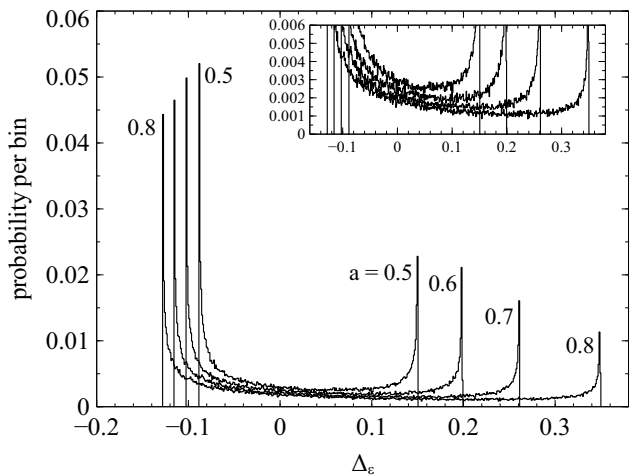}} \par}
    \caption{The simulated distributions of the correction $\Delta_{\varepsilon}$ (\ref{eq:delta_eps}) assuming
a simple cyclic process for $\delta \dot \Omega/\dot \Omega = \varepsilon =
a\cos\varphi$. The distributions are calculated for $a = 0.5, 0.6, 0.7$ and $0.8$, while the cyclic phase $\varphi$ is assumed to
be random over the interval $0-2\upi$. The sharp peaks at the boundary values
are caused by the harmonic nature of $\varepsilon$ (the cyclic variable spends more
time at its extrema) and the logarithmic nature of the relationship $\Delta_{\varepsilon}(\varepsilon)$.
See Section~\ref{sect:timingnoise} for details.}
    \label{fig:delta_epsilon}
\end{figure}

The extensive analysis of pulsar complex rotation undertaken by
\cite{hobbs10} showed that $\delta\dot\Omega$ is unlikely to be dominated by a
stochastic process. Instead, the observed timing noise patterns show
nearly periodic features. If this is the case, then the amplitude of
relative variations $\varepsilon$ can be estimated empirically from the statistics
of the observed second derivatives of pulsar spin frequencies $\ddot \Omega_{\rm obs} = \ddot \Omega + \delta \ddot \Omega$ , assuming them to be almost completely dominated by
the $\delta\ddot\Omega$ term. The corresponding analysis has been undertaken by
\cite{bbk12} and has shown that
\begin{equation}
	0.5 \la \max \vert \varepsilon \vert \la 0.8
	\label{eq:epsilon_constr}
\end{equation}
is able to reproduce the observed correlations between the pulsar timing parameters. Such amplitudes may lead to very high, up
to $\sim 0.35$, values of the $\Delta_{\varepsilon}$ correction, and are also slightly asymmetric with respect to $\Delta_{\varepsilon} = 0$. We plot the possible distributions
of this quantity in Fig. ~\ref{fig:delta_epsilon}, assuming $\varepsilon = a\cos\varphi$ for various $a$ and
random $\varphi$, as has been done by \citet{bbk12}. Thus, the term $\Delta_{\varepsilon}$ can, in principle, be a major source of the uncertainty in
the timing-based magnetic field estimation.

On the other hand, there are some indirect arguments in favour of
the idea that the pulsar secular spin-down follows the law (\ref{eq:spindown_general}) with
small typical values of $\varepsilon$. Thus, it has been shown many times that the
observable parameters of Galactic pulsars can be well reproduced
within a population synthesis adopting (\ref{eq:spindown_general}) with various forms of
$f(\alpha)$ \citep
[e.g][]{fgk06, ridley10, gullon14} and neglecting the corrections due to $\varepsilon \neq 0$.
Moreover, the model-independent analysis of pulsar kinetics on the
$P-\dot P$ diagram resulted in a reasonable value of their birthrate of
$\la 1$ century$^{-1}$, simply assuming $f(\alpha) \equiv 1$ \citep{kk08, vm11} and $\varepsilon \equiv 0$.

Finally, it basically remains unclear whether the correction $\Delta_{\varepsilon}$
should really be taken into account, because the value of the fraction
$r_{\rm ext}$ remains completely unknown.

\subsection{Astrophysical implications}
\label{sect:implicat}

Accurate and precise measurements of NS surface magnetic fields
are important in many areas, in particular for the investigation of
the field evolution $B(t)$ for normal, rotation-powered pulsars. The
NS magnetic field decay has been predicted in many theoretical
works \citep{ogunn69, goldr92, cumming04, pons07,
geppert09, vigano13, ip15}. On the other hand,
simulations of the observed pulsar population by \cite{fgk06} have shown that the statistics of basic observed
pulsar parameters can be well reproduced without the assumption
of the evolution of the magnetic field. This contradiction can in
principle be solved by the direct probing of the magnetic field
evolution of normal pulsars.

Indeed, it is expected that the surface $B$ strength decreases
down to one order of magnitude during the pulsar lifetime. This
is much greater than the typical uncertainty of the refined estimator $\log B_{\rm md} + \Delta^{\rm (eos)}_{\rm B}$, and three times greater if the maximal
$\Delta_{\varepsilon}$ is adopted. At the same time, the average correction $\langle \Delta^{\rm 
(eos)}_{\rm B}\rangle$ remains
common for all pulsars within a given EOS. So, the shape of an apparent field evolution $\log B^{\rm (eos)}(t)$ does not depend on the choice
of the specific EOS. This, however, is not true for the average value
of the initial field $\log
B^{\rm (eos)}(0)$.

We will present the results of our research into the empirical
magnetic field evolution for normal radiopulsars with independently
measured ages in a forthcoming paper.

The precise magnetic field measurements can also be useful when
the specific threshold value of $B$ exists within a problem. Thus,
there are 16 high-B isolated normal pulsars in the ATNF data base\footnote{\url{http://www.atnf.csiro.au/research/pulsar/psrcat/},
\cite{atnf}}
with the standard magnetic field $B_{\rm md}$ at the pole greater than the
Schwinger quantum limit of $\approx 4.4\times
10^{13}$ G. However, according to
the refined $B$ estimator we propose that there is only one (for BSK21)
or even zero (for MS1) such extreme objects at the 99.7 per cent
confidence level assuming $r_{\rm ext} = 0$.

Finally, the so-called pulsar `deathline' depends on both the current value of the magnetic field and on the spin period of a radiopulsar \citep[e.g][]{rs75, chenrud93, kantor04}. Thus, the population of `zombie'-pulsars
can be investigated within a given EOS, if the corrected value of $B$
is adopted.

\section{Conclusions}
\label{sect:conclude}

The basic results of our research are as follows:
\begin{enumerate}
\item The refined version of the canonical timing-based estimator of
the surface magnetic field of normal radiopulsars $B_{\rm md}(P,
\dot P)$ was introduced in a form $\log B = \log B_{\rm md} +
\Delta^{\rm (eos)}_{\rm B}$. The correction $\Delta^{\rm
(eos)}_{\rm B}$ is generally dependent on the given EOS, the NS mass $M$ and the obliquity $\alpha$.

\item It was found
that within existing observational constraints on masses $M$ and obliquities $\alpha$ of isolated
radiopulsars, the value of this correction is distributed almost normally with the standard
deviation as small as $\approx 0.06..0.09$ dex for most of realistic
EOSs. The average value of $\Delta^{\rm (eos)}_{\rm
B}$ is, however, non-zero and covers the range from $\approx -0.55$
to $\approx -0.25$, depending on the choice of EOS.

\item The
generalized timing-based estimator $\log B^* = \log B_{\rm md} -
0.37 \pm 0.10$ was also introduced under the assumption of equal
chances for all 22 considered EOSs to be realized in
nature. It indicates that within the Spitkovsky's spin-down Law (\ref{eq:spitkovsky})
the magnetic field of an arbitrary radiopulsar can be measured with
$\lesssim 30\%$ relative error using the timing parameters only.

\item The pulsar timing noise can, in principle, affect the timing-based measurement
of pulsar magnetic field, so additional correction term $r_{\rm ext}\Delta_{\varepsilon}$
has to be taken into account. While the value of $\Delta_{\varepsilon}$ can be as large as $\sim 0.35$,
the fraction of the timing noise due to external physical processes (relative to the secular spin-down law (\ref{eq:spitkovsky})) 
$r_{\rm ext} \in 0..1$ remains completely unknown.

\end{enumerate}

\section*{Acknowledgements}
We are grateful to Sergey Karpov, who kindly read the manuscript
and made a number of valuable suggestions that improved its clarity.
The work was performed according to the Russian Government
Program of Competitive Growth of Kazan Federal University. Data
analysis and simulations used hardware and software supported by
the Russian Science Foundation grant No. 14-50-00043. Artyom
Astashenok thanks the Russian Ministry of Education and Science
for support (project 2058/60).



\bibliographystyle{mnras}

\begin{thebibliography}{}
\makeatletter
\relax
\def\mn@urlcharsother{\let\do\@makeother \do\$\do\&\do\#\do\^\do\_\do\%\do\~}
\def\mn@doi{\begingroup\mn@urlcharsother \@ifnextchar [ {\mn@doi@}
  {\mn@doi@[]}}
\def\mn@doi@[#1]#2{\def\@tempa{#1}\ifx\@tempa\@empty \href
  {http://dx.doi.org/#2} {doi:#2}\else \href {http://dx.doi.org/#2} {#1}\fi
  \endgroup}
\def\mn@eprint#1#2{\mn@eprint@#1:#2::\@nil}
\def\mn@eprint@arXiv#1{\href {http://arxiv.org/abs/#1} {{\tt arXiv:#1}}}
\def\mn@eprint@dblp#1{\href {http://dblp.uni-trier.de/rec/bibtex/#1.xml}
  {dblp:#1}}
\def\mn@eprint@#1:#2:#3:#4\@nil{\def\@tempa {#1}\def\@tempb {#2}\def\@tempc
  {#3}\ifx \@tempc \@empty \let \@tempc \@tempb \let \@tempb \@tempa \fi \ifx
  \@tempb \@empty \def\@tempb {arXiv}\fi \@ifundefined
  {mn@eprint@\@tempb}{\@tempb:\@tempc}{\expandafter \expandafter \csname
  mn@eprint@\@tempb\endcsname \expandafter{\@tempc}}}

\bibitem[\protect\citeauthoryear{{Akmal}, {Pandharipande}  \&
  {Ravenhall}}{{Akmal} et~al.}{1998}]{APR}
{Akmal} A.,  {Pandharipande} V.~R.,   {Ravenhall} D.~G.,  1998, \mn@doi [\prc]
  {10.1103/PhysRevC.58.1804}, \href
  {http://adsabs.harvard.edu/abs/1998PhRvC..58.1804A} {58, 1804}

\bibitem[\protect\citeauthoryear{{Alford}, {Braby}, {Paris}  \&
  {Reddy}}{{Alford} et~al.}{2005}]{ALF2}
{Alford} M.,  {Braby} M.,  {Paris} M.,   {Reddy} S.,  2005, \mn@doi [\apj]
  {10.1086/430902}, \href {http://adsabs.harvard.edu/abs/2005ApJ...629..969A}
  {629, 969}

\bibitem[\protect\citeauthoryear{{Antoniadis}}{{Antoniadis}}{2013}]{ant_thesis}
{Antoniadis} J.~I.,  2013, PhD thesis, University of Bonn

\bibitem[\protect\citeauthoryear{{Antoniadis} et~al.,}{{Antoniadis}
  et~al.}{2013}]{Antoniadis}
{Antoniadis} J.,  et~al., 2013, \mn@doi [Science] {10.1126/science.1233232},
  \href {http://adsabs.harvard.edu/abs/2013Sci...340..448A} {340, 448}

\bibitem[\protect\citeauthoryear{{Antoniadis}, {Tauris}, {{\"O}zel}, {Barr},
  {Champion}  \& {Freire}}{{Antoniadis} et~al.}{2016}]{anto16}
{Antoniadis} J.,  {Tauris} T.~M.,  {{\"O}zel} F.,  {Barr} E.,  {Champion}
  D.~J.,   {Freire} P.~C.~C.,  2016, preprint, \href
  {http://adsabs.harvard.edu/abs/2016arXiv160501665A} {} (\mn@eprint {arXiv}
  {1605.01665})

\bibitem[\protect\citeauthoryear{{Archibald} et~al.,}{{Archibald}
  et~al.}{2016}]{archi16}
{Archibald} R.~F.,  et~al., 2016, \mn@doi [\apjl]
  {10.3847/2041-8205/819/1/L16}, \href
  {http://adsabs.harvard.edu/abs/2016ApJ...819L..16A} {819, L16}

\bibitem[\protect\citeauthoryear{{Arzamasskiy}, {Philippov}  \&
  {Tchekhovskoy}}{{Arzamasskiy} et~al.}{2015}]{arz15}
{Arzamasskiy} L.,  {Philippov} A.,   {Tchekhovskoy} A.,  2015, \mn@doi [\mnras]
  {10.1093/mnras/stv1818}, \href
  {http://adsabs.harvard.edu/abs/2015MNRAS.453.3540A} {453, 3540}

\bibitem[\protect\citeauthoryear{{Arzamasskiy}, {Beskin}  \&
  {Pirov}}{{Arzamasskiy} et~al.}{2017}]{arz16}
{Arzamasskiy} L.~I.,  {Beskin} V.~S.,   {Pirov} K.~K.,  2017, \mn@doi [\mnras]
  {10.1093/mnras/stw3139}, \href
  {http://adsabs.harvard.edu/abs/2017MNRAS.466.2325A} {466, 2325}

\bibitem[\protect\citeauthoryear{{Banik}, {Hempel}  \& {Bandyopadhyay}}{{Banik}
  et~al.}{2014}]{BHB}
{Banik} S.,  {Hempel} M.,   {Bandyopadhyay} D.,  2014, \mn@doi [\apjs]
  {10.1088/0067-0049/214/2/22}, \href
  {http://adsabs.harvard.edu/abs/2014ApJS..214...22B} {214, 22}

\bibitem[\protect\citeauthoryear{{Barsukov} \& {Tsygan}}{{Barsukov} \&
  {Tsygan}}{2010}]{bars10}
{Barsukov} D.~P.,  {Tsygan} A.~I.,  2010, \mn@doi [\mnras]
  {10.1111/j.1365-2966.2010.17365.x}, \href
  {http://adsabs.harvard.edu/abs/2010MNRAS.409.1077B} {409, 1077}

\bibitem[\protect\citeauthoryear{{Barsukov}, {Polyakova}  \&
  {Tsygan}}{{Barsukov} et~al.}{2009}]{bars09}
{Barsukov} D.~P.,  {Polyakova} P.~I.,   {Tsygan} A.~I.,  2009, \mn@doi
  [Astronomy Reports] {10.1134/S1063772909120075}, \href
  {http://adsabs.harvard.edu/abs/2009ARep...53.1146B} {53, 1146}

\bibitem[\protect\citeauthoryear{{Beskin}}{{Beskin}}{2016}]{bes16}
{Beskin} V.~S.,  2016, preprint, \href
  {http://adsabs.harvard.edu/abs/2016arXiv161003365B} {} (\mn@eprint {arXiv}
  {1610.03365})

\bibitem[\protect\citeauthoryear{{Beskin} \& {Nokhrina}}{{Beskin} \&
  {Nokhrina}}{2007}]{bes07}
{Beskin} V.~S.,  {Nokhrina} E.~E.,  2007, \mn@doi [\apss]
  {10.1007/s10509-007-9307-0}, \href
  {http://adsabs.harvard.edu/abs/2007Ap%26SS.308..569B} {308, 569}

\bibitem[\protect\citeauthoryear{{Beskin}, {Gurevich}  \& {Istomin}}{{Beskin}
  et~al.}{1993}]{bes93}
{Beskin} V.~S.,  {Gurevich} A.~V.,   {Istomin} Y.~N.,  1993, {Physics of the
  pulsar magnetosphere}.
{Cambridge, New York: Cambridge University Press}

\bibitem[\protect\citeauthoryear{{Beskin}, {Istomin}  \& {Philippov}}{{Beskin}
  et~al.}{2013}]{bes13}
{Beskin} V.~S.,  {Istomin} Y.~N.,   {Philippov} A.~A.,  2013, \mn@doi [Physics
  Uspekhi] {10.3367/UFNe.0183.201302e.0179}, \href
  {http://adsabs.harvard.edu/abs/2013PhyU...56..164B} {56, 164}

\bibitem[\protect\citeauthoryear{{Biryukov}, {Beskin}, {Karpov}  \&
  {Chmyreva}}{{Biryukov} et~al.}{2007}]{bbk07}
{Biryukov} A.,  {Beskin} G.,  {Karpov} S.,   {Chmyreva} L.,  2007, \mn@doi
  [Advances in Space Research] {10.1016/j.asr.2007.06.051}, \href
  {http://adsabs.harvard.edu/abs/2007AdSpR..40.1498B} {40, 1498}

\bibitem[\protect\citeauthoryear{{Biryukov}, {Beskin}  \& {Karpov}}{{Biryukov}
  et~al.}{2012}]{bbk12}
{Biryukov} A.,  {Beskin} G.,   {Karpov} S.,  2012, \mn@doi [\mnras]
  {10.1111/j.1365-2966.2011.20005.x}, \href
  {http://adsabs.harvard.edu/abs/2012MNRAS.420..103B} {420, 103}

\bibitem[\protect\citeauthoryear{{Borghese}, {Rea}, {Coti Zelati}, {Tiengo}  \&
  {Turolla}}{{Borghese} et~al.}{2015}]{borg15}
{Borghese} A.,  {Rea} N.,  {Coti Zelati} F.,  {Tiengo} A.,   {Turolla} R.,
  2015, \mn@doi [\apjl] {10.1088/2041-8205/807/1/L20}, \href
  {http://adsabs.harvard.edu/abs/2015ApJ...807L..20B} {807, L20}

\bibitem[\protect\citeauthoryear{{Boynton}, {Groth}, {Hutchinson}, {Nanos},
  {Partridge}  \& {Wilkinson}}{{Boynton} et~al.}{1972}]{boyn72}
{Boynton} P.~E.,  {Groth} E.~J.,  {Hutchinson} D.~P.,  {Nanos} Jr. G.~P.,
  {Partridge} R.~B.,   {Wilkinson} D.~T.,  1972, \mn@doi [\apj]
  {10.1086/151550}, \href {http://adsabs.harvard.edu/abs/1972ApJ...175..217B}
  {175, 217}

\bibitem[\protect\citeauthoryear{{Chabanat}, {Bonche}, {Haensel}, {Meyer}  \&
  {Schaeffer}}{{Chabanat} et~al.}{1998}]{SLy}
{Chabanat} E.,  {Bonche} P.,  {Haensel} P.,  {Meyer} J.,   {Schaeffer} R.,
  1998, \mn@doi [Nuclear Physics A] {10.1016/S0375-9474(98)00180-8}, \href
  {http://adsabs.harvard.edu/abs/1998NuPhA.635..231C} {635, 231}

\bibitem[\protect\citeauthoryear{{Chashkina} \& {Popov}}{{Chashkina} \&
  {Popov}}{2012}]{anna12}
{Chashkina} A.,  {Popov} S.~B.,  2012, \mn@doi [\na]
  {10.1016/j.newast.2012.01.004}, \href
  {http://adsabs.harvard.edu/abs/2012NewA...17..594C} {17, 594}

\bibitem[\protect\citeauthoryear{{Chen} \& {Ruderman}}{{Chen} \&
  {Ruderman}}{1993}]{chenrud93}
{Chen} K.,  {Ruderman} M.,  1993, \mn@doi [\apj] {10.1086/172129}, \href
  {http://adsabs.harvard.edu/abs/1993ApJ...402..264C} {402, 264}

\bibitem[\protect\citeauthoryear{{Cheng}}{{Cheng}}{1987}]{cheng87}
{Cheng} K.~S.,  1987, \mn@doi [\apj] {10.1086/165672}, \href
  {http://adsabs.harvard.edu/abs/1987ApJ...321..799C} {321, 799}

\bibitem[\protect\citeauthoryear{{Clark}, {Goodwin}, {Crowther}, {Kaper},
  {Fairbairn}, {Langer}  \& {Brocksopp}}{{Clark} et~al.}{2002}]{clark02}
{Clark} J.~S.,  {Goodwin} S.~P.,  {Crowther} P.~A.,  {Kaper} L.,  {Fairbairn}
  M.,  {Langer} N.,   {Brocksopp} C.,  2002, \mn@doi [\aap]
  {10.1051/0004-6361:20021184}, \href
  {http://adsabs.harvard.edu/abs/2002A&A...392..909C} {392, 909}

\bibitem[\protect\citeauthoryear{{Contopoulos}}{{Contopoulos}}{2007}]{cont07}
{Contopoulos} I.,  2007, \mn@doi [\aap] {10.1051/0004-6361:20078108}, \href
  {http://adsabs.harvard.edu/abs/2007A%26A...475..639C} {475, 639}

\bibitem[\protect\citeauthoryear{{Contopoulos} \& {Spitkovsky}}{{Contopoulos}
  \& {Spitkovsky}}{2006}]{cs06}
{Contopoulos} I.,  {Spitkovsky} A.,  2006, \mn@doi [\apj] {10.1086/501161},
  \href {http://adsabs.harvard.edu/abs/2006ApJ...643.1139C} {643, 1139}

\bibitem[\protect\citeauthoryear{{Cordes} \& {Helfand}}{{Cordes} \&
  {Helfand}}{1980}]{ch80}
{Cordes} J.~M.,  {Helfand} D.~J.,  1980, \mn@doi [\apj] {10.1086/158150}, \href
  {http://adsabs.harvard.edu/abs/1980ApJ...239..640C} {239, 640}

\bibitem[\protect\citeauthoryear{{Cumming}, {Arras}  \& {Zweibel}}{{Cumming}
  et~al.}{2004}]{cumming04}
{Cumming} A.,  {Arras} P.,   {Zweibel} E.,  2004, \mn@doi [\apj]
  {10.1086/421324}, \href {http://adsabs.harvard.edu/abs/2004ApJ...609..999C}
  {609, 999}

\bibitem[\protect\citeauthoryear{{D'Alessandro}, {McCulloch}, {Hamilton}  \&
  {Deshpande}}{{D'Alessandro} et~al.}{1995}]{daless95}
{D'Alessandro} F.,  {McCulloch} P.~M.,  {Hamilton} P.~A.,   {Deshpande} A.~A.,
  1995, \mn@doi [\mnras] {10.1093/mnras/277.3.1033}, \href
  {http://adsabs.harvard.edu/abs/1995MNRAS.277.1033D} {277, 1033}

\bibitem[\protect\citeauthoryear{{Danielewicz} \& {Lee}}{{Danielewicz} \&
  {Lee}}{2009}]{SK16-2}
{Danielewicz} P.,  {Lee} J.,  2009, \mn@doi [Nuclear Physics A]
  {10.1016/j.nuclphysa.2008.11.007}, \href
  {http://adsabs.harvard.edu/abs/2009NuPhA.818...36D} {818, 36}

\bibitem[\protect\citeauthoryear{{Davis} \& {Goldstein}}{{Davis} \&
  {Goldstein}}{1970}]{dg70}
{Davis} L.,  {Goldstein} M.,  1970, \mn@doi [\apjl] {10.1086/180482}, \href
  {http://adsabs.harvard.edu/abs/1970ApJ...159L..81D} {159}

\bibitem[\protect\citeauthoryear{{Demorest}, {Pennucci}, {Ransom}, {Roberts}
  \& {Hessels}}{{Demorest} et~al.}{2010}]{Demorest}
{Demorest} P.~B.,  {Pennucci} T.,  {Ransom} S.~M.,  {Roberts} M.~S.~E.,
  {Hessels} J.~W.~T.,  2010, \mn@doi [\nat] {10.1038/nature09466}, \href
  {http://adsabs.harvard.edu/abs/2010Natur.467.1081D} {467, 1081}

\bibitem[\protect\citeauthoryear{{Deutsch}}{{Deutsch}}{1955}]{deu55}
{Deutsch} A.~J.,  1955, Annales d'Astrophysique, \href
  {http://adsabs.harvard.edu/abs/1955AnAp...18....1D} {18, 1}

\bibitem[\protect\citeauthoryear{{Dobaczewski}, {Nazarewicz}, {Werner},
  {Berger}, {Chinn}  \& {Decharg{\'e}}}{{Dobaczewski} et~al.}{1996}]{SK16-1}
{Dobaczewski} J.,  {Nazarewicz} W.,  {Werner} T.~R.,  {Berger} J.~F.,  {Chinn}
  C.~R.,   {Decharg{\'e}} J.,  1996, \mn@doi [\prc] {10.1103/PhysRevC.53.2809},
  \href {http://adsabs.harvard.edu/abs/1996PhRvC..53.2809D} {53, 2809}

\bibitem[\protect\citeauthoryear{{Douchin} \& {Haensel}}{{Douchin} \&
  {Haensel}}{2001}]{SLy-4}
{Douchin} F.,  {Haensel} P.,  2001, \mn@doi [\aap]
  {10.1051/0004-6361:20011402}, \href
  {http://adsabs.harvard.edu/abs/2001A%26A...380..151D} {380, 151}

\bibitem[\protect\citeauthoryear{{Ek{\c s}i}, {Anda{\c c}}, {{\c
  C}{\i}k{\i}nto{\u g}lu}, {G{\"u}gercino{\u g}lu}, {Vahdat Motlagh}  \&
  {K{\i}z{\i}ltan}}{{Ek{\c s}i} et~al.}{2016}]{eksi16}
{Ek{\c s}i} K.~Y.,  {Anda{\c c}} I.~C.,  {{\c C}{\i}k{\i}nto{\u g}lu} S.,
  {G{\"u}gercino{\u g}lu} E.,  {Vahdat Motlagh} A.,   {K{\i}z{\i}ltan} B.,
  2016, \mn@doi [\apj] {10.3847/0004-637X/823/1/34}, \href
  {http://adsabs.harvard.edu/abs/2016ApJ...823...34E} {823, 34}

\bibitem[\protect\citeauthoryear{{Engvik}, {Hjorth-Jensen}, {Osnes}, {Bao}  \&
  {{\O}stgaard}}{{Engvik} et~al.}{1994}]{ENG}
{Engvik} L.,  {Hjorth-Jensen} M.,  {Osnes} E.,  {Bao} G.,   {{\O}stgaard} E.,
  1994, \mn@doi [Physical Review Letters] {10.1103/PhysRevLett.73.2650}, \href
  {http://adsabs.harvard.edu/abs/1994PhRvL..73.2650E} {73, 2650}

\bibitem[\protect\citeauthoryear{{Faucher-Gigu{\`e}re} \&
  {Kaspi}}{{Faucher-Gigu{\`e}re} \& {Kaspi}}{2006}]{fgk06}
{Faucher-Gigu{\`e}re} C.-A.,  {Kaspi} V.~M.,  2006, \mn@doi [\apj]
  {10.1086/501516}, \href {http://adsabs.harvard.edu/abs/2006ApJ...643..332F}
  {643, 332}

\bibitem[\protect\citeauthoryear{{Finn}}{{Finn}}{1994}]{finn94}
{Finn} L.~S.,  1994, \mn@doi [Physical Review Letters]
  {10.1103/PhysRevLett.73.1878}, \href
  {http://adsabs.harvard.edu/abs/1994PhRvL..73.1878F} {73, 1878}

\bibitem[\protect\citeauthoryear{{Fischer} et~al.,}{{Fischer}
  et~al.}{2011}]{HSHEN-3}
{Fischer} T.,  et~al., 2011, \mn@doi [\apjs] {10.1088/0067-0049/194/2/39},
  \href {http://adsabs.harvard.edu/abs/2011ApJS..194...39F} {194, 39}

\bibitem[\protect\citeauthoryear{{Gaitanos}, {Di Toro}, {Typel}, {Baran},
  {Fuchs}, {Greco}  \& {Wolter}}{{Gaitanos} et~al.}{2004}]{DDH}
{Gaitanos} T.,  {Di Toro} M.,  {Typel} S.,  {Baran} V.,  {Fuchs} C.,  {Greco}
  V.,   {Wolter} H.~H.,  2004, \mn@doi [Nuclear Physics A]
  {10.1016/j.nuclphysa.2003.12.001}, \href
  {http://adsabs.harvard.edu/abs/2004NuPhA.732...24G} {732, 24}

\bibitem[\protect\citeauthoryear{{Geppert}}{{Geppert}}{2009}]{geppert09}
{Geppert} U.,  2009, in {Becker} W.,  ed.,  Vol. 357, Astrophysics and Space
  Science Library. p.~319 (\mn@eprint {} {astro-ph/0611708})

\bibitem[\protect\citeauthoryear{{Glendenning}}{{Glendenning}}{1985}]{GNH}
{Glendenning} N.~K.,  1985, \mn@doi [\apj] {10.1086/163253}, \href
  {http://adsabs.harvard.edu/abs/1985ApJ...293..470G} {293, 470}

\bibitem[\protect\citeauthoryear{{Glendenning} \& {Moszkowski}}{{Glendenning}
  \& {Moszkowski}}{1991}]{GM}
{Glendenning} N.~K.,  {Moszkowski} S.~A.,  1991, \mn@doi [Physical Review
  Letters] {10.1103/PhysRevLett.67.2414}, \href
  {http://adsabs.harvard.edu/abs/1991PhRvL..67.2414G} {67, 2414}

\bibitem[\protect\citeauthoryear{{Goldreich} \& {Julian}}{{Goldreich} \&
  {Julian}}{1969}]{gj69}
{Goldreich} P.,  {Julian} W.~H.,  1969, \mn@doi [\apj] {10.1086/150119}, \href
  {http://adsabs.harvard.edu/abs/1969ApJ...157..869G} {157, 869}

\bibitem[\protect\citeauthoryear{{Goldreich} \& {Reisenegger}}{{Goldreich} \&
  {Reisenegger}}{1992}]{goldr92}
{Goldreich} P.,  {Reisenegger} A.,  1992, \mn@doi [\apj] {10.1086/171646},
  \href {http://adsabs.harvard.edu/abs/1992ApJ...395..250G} {395, 250}

\bibitem[\protect\citeauthoryear{{Good} \& {Ng}}{{Good} \& {Ng}}{1985}]{good85}
{Good} M.~L.,  {Ng} K.~K.,  1985, \mn@doi [\apj] {10.1086/163736}, \href
  {http://adsabs.harvard.edu/abs/1985ApJ...299..706G} {299, 706}

\bibitem[\protect\citeauthoryear{{Goriely}, {Chamel}  \& {Pearson}}{{Goriely}
  et~al.}{2010}]{BSKEOS}
{Goriely} S.,  {Chamel} N.,   {Pearson} J.~M.,  2010, \mn@doi [\prc]
  {10.1103/PhysRevC.82.035804}, \href
  {http://adsabs.harvard.edu/abs/2010PhRvC..82c5804G} {82, 035804}

\bibitem[\protect\citeauthoryear{{Gould}}{{Gould}}{1994}]{gould94}
{Gould} D.~M.,  1994, PhD thesis, Univ.~of Manchester

\bibitem[\protect\citeauthoryear{{Grill}, {Pais}, {Provid{\^e}ncia},
  {Vida{\~n}a}  \& {Avancini}}{{Grill} et~al.}{2014}]{DDH-2}
{Grill} F.,  {Pais} H.,  {Provid{\^e}ncia} C.,  {Vida{\~n}a} I.,   {Avancini}
  S.~S.,  2014, \mn@doi [\prc] {10.1103/PhysRevC.90.045803}, \href
  {http://adsabs.harvard.edu/abs/2014PhRvC..90d5803G} {90, 045803}

\bibitem[\protect\citeauthoryear{{Gull{\'o}n}, {Miralles}, {Vigan{\`o}}  \&
  {Pons}}{{Gull{\'o}n} et~al.}{2014}]{gullon14}
{Gull{\'o}n} M.,  {Miralles} J.~A.,  {Vigan{\`o}} D.,   {Pons} J.~A.,  2014,
  \mn@doi [\mnras] {10.1093/mnras/stu1253}, \href
  {http://adsabs.harvard.edu/abs/2014MNRAS.443.1891G} {443, 1891}

\bibitem[\protect\citeauthoryear{{Gulminelli} \& {Raduta}}{{Gulminelli} \&
  {Raduta}}{2015}]{SK16}
{Gulminelli} F.,  {Raduta} A.~R.,  2015, \mn@doi [\prc]
  {10.1103/PhysRevC.92.055803}, \href
  {http://adsabs.harvard.edu/abs/2015PhRvC..92e5803G} {92, 055803}

\bibitem[\protect\citeauthoryear{{Hambaryan}, {Suleimanov}, {Schwope},
  {Neuh{\"a}user}, {Werner}  \& {Potekhin}}{{Hambaryan} et~al.}{2011}]{HAMB}
{Hambaryan} V.,  {Suleimanov} V.,  {Schwope} A.~D.,  {Neuh{\"a}user} R.,
  {Werner} K.,   {Potekhin} A.~Y.,  2011, \mn@doi [\aap]
  {10.1051/0004-6361/201117548}, \href
  {http://adsabs.harvard.edu/abs/2011A%26A...534A..74H} {534, A74}

\bibitem[\protect\citeauthoryear{{Hamil}, {Stone}, {Urbanec}  \&
  {Urbancov{\'a}}}{{Hamil} et~al.}{2015}]{hamil15}
{Hamil} O.,  {Stone} J.~R.,  {Urbanec} M.,   {Urbancov{\'a}} G.,  2015, \mn@doi
  [\prd] {10.1103/PhysRevD.91.063007}, \href
  {http://adsabs.harvard.edu/abs/2015PhRvD..91f3007H} {91, 063007}

\bibitem[\protect\citeauthoryear{{Hamil}, {Stone}  \& {Stone}}{{Hamil}
  et~al.}{2016}]{hamil16}
{Hamil} O.,  {Stone} N.~J.,   {Stone} J.~R.,  2016, \mn@doi [\prd]
  {10.1103/PhysRevD.94.063012}, \href
  {http://adsabs.harvard.edu/abs/2016PhRvD..94f3012H} {94, 063012}

\bibitem[\protect\citeauthoryear{{Hempel} \& {Schaffner-Bielich}}{{Hempel} \&
  {Schaffner-Bielich}}{2010}]{BHB-2}
{Hempel} M.,  {Schaffner-Bielich} J.,  2010, \mn@doi [Nuclear Physics A]
  {10.1016/j.nuclphysa.2010.02.010}, \href
  {http://adsabs.harvard.edu/abs/2010NuPhA.837..210H} {837, 210}

\bibitem[\protect\citeauthoryear{{Hobbs}, {Lyne}, {Kramer}, {Martin}  \&
  {Jordan}}{{Hobbs} et~al.}{2004}]{hobbs04}
{Hobbs} G.,  {Lyne} A.~G.,  {Kramer} M.,  {Martin} C.~E.,   {Jordan} C.,  2004,
  \mn@doi [\mnras] {10.1111/j.1365-2966.2004.08157.x}, \href
  {http://adsabs.harvard.edu/abs/2004MNRAS.353.1311H} {353, 1311}

\bibitem[\protect\citeauthoryear{{Hobbs}, {Lyne}  \& {Kramer}}{{Hobbs}
  et~al.}{2010}]{hobbs10}
{Hobbs} G.,  {Lyne} A.~G.,   {Kramer} M.,  2010, \mn@doi [\mnras]
  {10.1111/j.1365-2966.2009.15938.x}, \href
  {http://adsabs.harvard.edu/abs/2010MNRAS.402.1027H} {402, 1027}

\bibitem[\protect\citeauthoryear{{Igoshev} \& {Popov}}{{Igoshev} \&
  {Popov}}{2015}]{ip15}
{Igoshev} A.~P.,  {Popov} S.~B.,  2015, \mn@doi [Astronomische Nachrichten]
  {10.1002/asna.201512232}, \href
  {http://adsabs.harvard.edu/abs/2015AN....336..831I} {336, 831}

\bibitem[\protect\citeauthoryear{{Janssen} \& {Stappers}}{{Janssen} \&
  {Stappers}}{2006}]{janssen06}
{Janssen} G.~H.,  {Stappers} B.~W.,  2006, \mn@doi [\aap]
  {10.1051/0004-6361:20065267}, \href
  {http://adsabs.harvard.edu/abs/2006A%26A...457..611J} {457, 611}

\bibitem[\protect\citeauthoryear{{Kalapotharakos} \&
  {Contopoulos}}{{Kalapotharakos} \& {Contopoulos}}{2009}]{kala09}
{Kalapotharakos} C.,  {Contopoulos} I.,  2009, \mn@doi [\aap]
  {10.1051/0004-6361:200810281}, \href
  {http://adsabs.harvard.edu/abs/2009A%26A...496..495K} {496, 495}

\bibitem[\protect\citeauthoryear{{Kantor} \& {Tsygan}}{{Kantor} \&
  {Tsygan}}{2004}]{kantor04}
{Kantor} E.~M.,  {Tsygan} A.~I.,  2004, \mn@doi [Astronomy Reports]
  {10.1134/1.1836026}, \href
  {http://adsabs.harvard.edu/abs/2004ARep...48.1029K} {48, 1029}

\bibitem[\protect\citeauthoryear{{Keane} \& {Kramer}}{{Keane} \&
  {Kramer}}{2008}]{kk08}
{Keane} E.~F.,  {Kramer} M.,  2008, \mn@doi [\mnras]
  {10.1111/j.1365-2966.2008.14045.x}, \href
  {http://adsabs.harvard.edu/abs/2008MNRAS.391.2009K} {391, 2009}

\bibitem[\protect\citeauthoryear{{Kiziltan}, {Kottas}, {De Yoreo}  \&
  {Thorsett}}{{Kiziltan} et~al.}{2013}]{kiz13}
{Kiziltan} B.,  {Kottas} A.,  {De Yoreo} M.,   {Thorsett} S.~E.,  2013, \mn@doi
  [\apj] {10.1088/0004-637X/778/1/66}, \href
  {http://adsabs.harvard.edu/abs/2013ApJ...778...66K} {778, 66}

\bibitem[\protect\citeauthoryear{{Kou} \& {Tong}}{{Kou} \&
  {Tong}}{2015}]{kou15}
{Kou} F.~F.,  {Tong} H.,  2015, \mn@doi [\mnras] {10.1093/mnras/stv734}, \href
  {http://adsabs.harvard.edu/abs/2015MNRAS.450.1990K} {450, 1990}

\bibitem[\protect\citeauthoryear{{Kou}, {Tong}  \& {Wang}}{{Kou}
  et~al.}{2016}]{kou16}
{Kou} F.~F.,  {Tong} H.,   {Wang} N.,  2016, preprint, \href
  {http://adsabs.harvard.edu/abs/2016arXiv160401231K} {} (\mn@eprint {arXiv}
  {1604.01231})

\bibitem[\protect\citeauthoryear{{Kramer}, {Lyne}, {O'Brien}, {Jordan}  \&
  {Lorimer}}{{Kramer} et~al.}{2006}]{kramer06}
{Kramer} M.,  {Lyne} A.~G.,  {O'Brien} J.~T.,  {Jordan} C.~A.,   {Lorimer}
  D.~R.,  2006, \mn@doi [Science] {10.1126/science.1124060}, \href
  {http://adsabs.harvard.edu/abs/2006Sci...312..549K} {312, 549}

\bibitem[\protect\citeauthoryear{{Lattimer}}{{Lattimer}}{2012}]{lat12}
{Lattimer} J.~M.,  2012, \mn@doi [Annual Review of Nuclear and Particle
  Science] {10.1146/annurev-nucl-102711-095018}, \href
  {http://adsabs.harvard.edu/abs/2012ARNPS..62..485L} {62, 485}

\bibitem[\protect\citeauthoryear{{Leahy}, {Morsink}  \& {Chou}}{{Leahy}
  et~al.}{2011}]{Leahy}
{Leahy} D.~A.,  {Morsink} S.~M.,   {Chou} Y.,  2011, \mn@doi [\apj]
  {10.1088/0004-637X/742/1/17}, \href
  {http://adsabs.harvard.edu/abs/2011ApJ...742...17L} {742, 17}

\bibitem[\protect\citeauthoryear{{Lyne} \& {Manchester}}{{Lyne} \&
  {Manchester}}{1988}]{lyne88}
{Lyne} A.~G.,  {Manchester} R.~N.,  1988, \mn@doi [\mnras]
  {10.1093/mnras/234.3.477}, \href
  {http://adsabs.harvard.edu/abs/1988MNRAS.234..477L} {234, 477}

\bibitem[\protect\citeauthoryear{{Lyne}, {Hobbs}, {Kramer}, {Stairs}  \&
  {Stappers}}{{Lyne} et~al.}{2010}]{lyne10}
{Lyne} A.,  {Hobbs} G.,  {Kramer} M.,  {Stairs} I.,   {Stappers} B.,  2010,
  \mn@doi [Science] {10.1126/science.1186683}, \href
  {http://adsabs.harvard.edu/abs/2010Sci...329..408L} {329, 408}

\bibitem[\protect\citeauthoryear{{Lyne}, {Graham-Smith}, {Weltevrede},
  {Jordan}, {Stappers}, {Bassa}  \& {Kramer}}{{Lyne} et~al.}{2013}]{lyne13}
{Lyne} A.,  {Graham-Smith} F.,  {Weltevrede} P.,  {Jordan} C.,  {Stappers} B.,
  {Bassa} C.,   {Kramer} M.,  2013, \mn@doi [Science]
  {10.1126/science.1243254}, \href
  {http://adsabs.harvard.edu/abs/2013Sci...342..598L} {342, 598}

\bibitem[\protect\citeauthoryear{{Malov}}{{Malov}}{2016}]{malov16}
{Malov} I.~F.,  2016, preprint, \href
  {http://adsabs.harvard.edu/abs/2016arXiv160808084M} {} (\mn@eprint {arXiv}
  {1608.08084})

\bibitem[\protect\citeauthoryear{{Malov} \& {Nikitina}}{{Malov} \&
  {Nikitina}}{2011}]{nikitina11a}
{Malov} I.~F.,  {Nikitina} E.~B.,  2011, \mn@doi [Astronomy Reports]
  {10.1134/S1063772911010021}, \href
  {http://adsabs.harvard.edu/abs/2011ARep...55...19M} {55, 19}

\bibitem[\protect\citeauthoryear{{Manchester} \& {Taylor}}{{Manchester} \&
  {Taylor}}{1977}]{mt77}
{Manchester} R.~N.,  {Taylor} J.~H.,  1977, {Pulsars}.
San Francisco : W.~H.~Freeman

\bibitem[\protect\citeauthoryear{{Manchester}, {Hobbs}, {Teoh}  \&
  {Hobbs}}{{Manchester} et~al.}{2005}]{atnf}
{Manchester} R.~N.,  {Hobbs} G.~B.,  {Teoh} A.,   {Hobbs} M.,  2005, \mn@doi
  [\aj] {10.1086/428488}, \href
  {http://adsabs.harvard.edu/abs/2005AJ....129.1993M} {129, 1993}

\bibitem[\protect\citeauthoryear{{Marshall}, {Guillemot}, {Harding}, {Martin}
  \& {Smith}}{{Marshall} et~al.}{2016}]{marsh16}
{Marshall} F.~E.,  {Guillemot} L.,  {Harding} A.~K.,  {Martin} P.,   {Smith}
  D.~A.,  2016, \mn@doi [\apjl] {10.3847/2041-8205/827/2/L39}, \href
  {http://adsabs.harvard.edu/abs/2016ApJ...827L..39M} {827, L39}

\bibitem[\protect\citeauthoryear{{Melatos}}{{Melatos}}{2000}]{melatos2000}
{Melatos} A.,  2000, \mn@doi [\mnras] {10.1046/j.1365-8711.2000.03031.x}, \href
  {http://adsabs.harvard.edu/abs/2000MNRAS.313..217M} {313, 217}

\bibitem[\protect\citeauthoryear{{Melatos} \& {Link}}{{Melatos} \&
  {Link}}{2014}]{ml14}
{Melatos} A.,  {Link} B.,  2014, \mn@doi [\mnras] {10.1093/mnras/stt1828},
  \href {http://adsabs.harvard.edu/abs/2014MNRAS.437...21M} {437, 21}

\bibitem[\protect\citeauthoryear{{Michel}}{{Michel}}{1973}]{michel73}
{Michel} F.~C.,  1973, \mn@doi [\apj] {10.1086/151956}, \href
  {http://adsabs.harvard.edu/abs/1973ApJ...180..207M} {180, 207}

\bibitem[\protect\citeauthoryear{{M{\"u}ller} \& {Serot}}{{M{\"u}ller} \&
  {Serot}}{1996}]{MS}
{M{\"u}ller} H.,  {Serot} B.~D.,  1996, \mn@doi [Nuclear Physics A]
  {10.1016/0375-9474(96)00187-X}, \href
  {http://adsabs.harvard.edu/abs/1996NuPhA.606..508M} {606, 508}

\bibitem[\protect\citeauthoryear{{M{\"u}ther}, {Prakash}  \&
  {Ainsworth}}{{M{\"u}ther} et~al.}{1987}]{MPA}
{M{\"u}ther} H.,  {Prakash} M.,   {Ainsworth} T.~L.,  1987, \mn@doi [Physics
  Letters B] {10.1016/0370-2693(87)91611-X}, \href
  {http://adsabs.harvard.edu/abs/1987PhLB..199..469M} {199, 469}

\bibitem[\protect\citeauthoryear{{Nayyar} \& {Owen}}{{Nayyar} \&
  {Owen}}{2006}]{H4}
{Nayyar} M.,  {Owen} B.~J.,  2006, \mn@doi [\prd] {10.1103/PhysRevD.73.084001},
  \href {http://adsabs.harvard.edu/abs/2006PhRvD..73h4001N} {73, 084001}

\bibitem[\protect\citeauthoryear{{Nice} et~al.,}{{Nice} et~al.}{2013}]{nice13}
{Nice} D.~J.,  et~al., 2013, \mn@doi [\apj] {10.1088/0004-637X/772/1/50}, \href
  {http://adsabs.harvard.edu/abs/2013ApJ...772...50N} {772, 50}

\bibitem[\protect\citeauthoryear{{Nikitina} \& {Malov}}{{Nikitina} \&
  {Malov}}{2016}]{nikitina16}
{Nikitina} E.~B.,  {Malov} I.~F.,  2016, preprint, \href
  {http://adsabs.harvard.edu/abs/2016arXiv160808525N} {} (\mn@eprint {arXiv}
  {1608.08525})

\bibitem[\protect\citeauthoryear{{Oertel}, {Provid{\^e}ncia}, {Gulminelli}  \&
  {Raduta}}{{Oertel} et~al.}{2015}]{GM-1}
{Oertel} M.,  {Provid{\^e}ncia} C.,  {Gulminelli} F.,   {Raduta} A.~R.,  2015,
  \mn@doi [Journal of Physics G Nuclear Physics]
  {10.1088/0954-3899/42/7/075202}, \href
  {http://adsabs.harvard.edu/abs/2015JPhG...42g5202O} {42, 075202}

\bibitem[\protect\citeauthoryear{{Ostriker} \& {Gunn}}{{Ostriker} \&
  {Gunn}}{1969}]{ogunn69}
{Ostriker} J.~P.,  {Gunn} J.~E.,  1969, \mn@doi [\apj] {10.1086/150160}, \href
  {http://adsabs.harvard.edu/abs/1969ApJ...157.1395O} {157, 1395}

\bibitem[\protect\citeauthoryear{{Ou}, {Tong}, {Kou}  \& {Ding}}{{Ou}
  et~al.}{2016}]{ou16}
{Ou} Z.~W.,  {Tong} H.,  {Kou} F.~F.,   {Ding} G.~Q.,  2016, \mn@doi [\mnras]
  {10.1093/mnras/stw227}, \href
  {http://adsabs.harvard.edu/abs/2016MNRAS.457.3922O} {457, 3922}

\bibitem[\protect\citeauthoryear{{{\"O}zel} \& {Freire}}{{{\"O}zel} \&
  {Freire}}{2016}]{ozel16}
{{\"O}zel} F.,  {Freire} P.,  2016, \mn@doi [\araa]
  {10.1146/annurev-astro-081915-023322}, \href
  {http://adsabs.harvard.edu/abs/2016ARA%26A..54..401O} {54, 401}

\bibitem[\protect\citeauthoryear{{{\"O}zel}, {Psaltis}, {Narayan}  \& {Santos
  Villarreal}}{{{\"O}zel} et~al.}{2012}]{ozel12}
{{\"O}zel} F.,  {Psaltis} D.,  {Narayan} R.,   {Santos Villarreal} A.,  2012,
  \mn@doi [\apj] {10.1088/0004-637X/757/1/55}, \href
  {http://adsabs.harvard.edu/abs/2012ApJ...757...55O} {757, 55}

\bibitem[\protect\citeauthoryear{{Pearson}, {Goriely}  \& {Chamel}}{{Pearson}
  et~al.}{2011}]{BSKEOS-1}
{Pearson} J.~M.,  {Goriely} S.,   {Chamel} N.,  2011, \mn@doi [\prc]
  {10.1103/PhysRevC.83.065810}, \href
  {http://adsabs.harvard.edu/abs/2011PhRvC..83f5810P} {83, 065810}

\bibitem[\protect\citeauthoryear{{Pearson}, {Chamel}, {Goriely}  \&
  {Ducoin}}{{Pearson} et~al.}{2012}]{BSKEOS-2}
{Pearson} J.~M.,  {Chamel} N.,  {Goriely} S.,   {Ducoin} C.,  2012, \mn@doi
  [\prc] {10.1103/PhysRevC.85.065803}, \href
  {http://adsabs.harvard.edu/abs/2012PhRvC..85f5803P} {85, 065803}

\bibitem[\protect\citeauthoryear{{P{\'e}tri}}{{P{\'e}tri}}{2012}]{petri12}
{P{\'e}tri} J.,  2012, \mn@doi [\mnras] {10.1111/j.1365-2966.2012.21238.x},
  \href {http://adsabs.harvard.edu/abs/2012MNRAS.424..605P} {424, 605}

\bibitem[\protect\citeauthoryear{{Petrova}}{{Petrova}}{2016}]{petr16}
{Petrova} S.~A.,  2016, preprint, \href
  {http://adsabs.harvard.edu/abs/2016arXiv160807998P} {} (\mn@eprint {arXiv}
  {1608.07998})

\bibitem[\protect\citeauthoryear{{Philippov}, {Tchekhovskoy}  \&
  {Li}}{{Philippov} et~al.}{2014}]{phil14}
{Philippov} A.,  {Tchekhovskoy} A.,   {Li} J.~G.,  2014, \mn@doi [\mnras]
  {10.1093/mnras/stu591}, \href
  {http://adsabs.harvard.edu/abs/2014MNRAS.441.1879P} {441, 1879}

\bibitem[\protect\citeauthoryear{{Pons}, {Link}, {Miralles}  \&
  {Geppert}}{{Pons} et~al.}{2007}]{pons07}
{Pons} J.~A.,  {Link} B.,  {Miralles} J.~A.,   {Geppert} U.,  2007, \mn@doi
  [Physical Review Letters] {10.1103/PhysRevLett.98.071101}, \href
  {http://adsabs.harvard.edu/abs/2007PhRvL..98g1101P} {98, 071101}

\bibitem[\protect\citeauthoryear{{Pons}, {Vigan{\`o}}  \& {Geppert}}{{Pons}
  et~al.}{2012}]{pons12}
{Pons} J.~A.,  {Vigan{\`o}} D.,   {Geppert} U.,  2012, \mn@doi [\aap]
  {10.1051/0004-6361/201220091}, \href
  {http://adsabs.harvard.edu/abs/2012A%26A...547A...9P} {547, A9}

\bibitem[\protect\citeauthoryear{{Postnov} \& {Yungelson}}{{Postnov} \&
  {Yungelson}}{2014}]{postnov14}
{Postnov} K.~A.,  {Yungelson} L.~R.,  2014, \mn@doi [Living Reviews in
  Relativity] {10.12942/lrr-2014-3}, \href
  {http://adsabs.harvard.edu/abs/2014LRR....17....3P} {17}

\bibitem[\protect\citeauthoryear{{Rankin}}{{Rankin}}{1993a}]{rankin93a}
{Rankin} J.~M.,  1993a, \mn@doi [\apjs] {10.1086/191758}, \href
  {http://adsabs.harvard.edu/abs/1993ApJS...85..145R} {85, 145}

\bibitem[\protect\citeauthoryear{{Rankin}}{{Rankin}}{1993b}]{rankin93b}
{Rankin} J.~M.,  1993b, \mn@doi [\apj] {10.1086/172361}, \href
  {http://adsabs.harvard.edu/abs/1993ApJ...405..285R} {405, 285}

\bibitem[\protect\citeauthoryear{{Revnivtsev} \& {Mereghetti}}{{Revnivtsev} \&
  {Mereghetti}}{2015}]{revmer15}
{Revnivtsev} M.,  {Mereghetti} S.,  2015, \mn@doi [\ssr]
  {10.1007/s11214-014-0123-x}, \href
  {http://adsabs.harvard.edu/abs/2015SSRv..191..293R} {191, 293}

\bibitem[\protect\citeauthoryear{{Ridley} \& {Lorimer}}{{Ridley} \&
  {Lorimer}}{2010}]{ridley10}
{Ridley} J.~P.,  {Lorimer} D.~R.,  2010, \mn@doi [\mnras]
  {10.1111/j.1365-2966.2010.16342.x}, \href
  {http://adsabs.harvard.edu/abs/2010MNRAS.404.1081R} {404, 1081}

\bibitem[\protect\citeauthoryear{{Ruderman} \& {Sutherland}}{{Ruderman} \&
  {Sutherland}}{1975}]{rs75}
{Ruderman} M.~A.,  {Sutherland} P.~G.,  1975, \mn@doi [\apj] {10.1086/153393},
  \href {http://adsabs.harvard.edu/abs/1975ApJ...196...51R} {196, 51}

\bibitem[\protect\citeauthoryear{{Sagert}, {Fischer}, {Hempel}, {Pagliara},
  {Schaffner-Bielich}, {Mezzacappa}, {Thielemann}  \&
  {Liebend{\"o}rfer}}{{Sagert} et~al.}{2009}]{HSHEN}
{Sagert} I.,  {Fischer} T.,  {Hempel} M.,  {Pagliara} G.,  {Schaffner-Bielich}
  J.,  {Mezzacappa} A.,  {Thielemann} F.-K.,   {Liebend{\"o}rfer} M.,  2009,
  \mn@doi [Physical Review Letters] {10.1103/PhysRevLett.102.081101}, \href
  {http://adsabs.harvard.edu/abs/2009PhRvL.102h1101S} {102, 081101}

\bibitem[\protect\citeauthoryear{{Sagert}, {Fischer}, {Hempel}, {Pagliara},
  {Schaffner-Bielich}, {Thielemann}  \& {Liebend{\"o}rfer}}{{Sagert}
  et~al.}{2010}]{HSHEN-2}
{Sagert} I.,  {Fischer} T.,  {Hempel} M.,  {Pagliara} G.,  {Schaffner-Bielich}
  J.,  {Thielemann} F.-K.,   {Liebend{\"o}rfer} M.,  2010, \mn@doi [Journal of
  Physics G Nuclear Physics] {10.1088/0954-3899/37/9/094064}, \href
  {http://adsabs.harvard.edu/abs/2010JPhG...37i4064S} {37, 094064}

\bibitem[\protect\citeauthoryear{{Schwab}, {Podsiadlowski}  \&
  {Rappaport}}{{Schwab} et~al.}{2010}]{swa10}
{Schwab} J.,  {Podsiadlowski} P.,   {Rappaport} S.,  2010, \mn@doi [\apj]
  {10.1088/0004-637X/719/1/722}, \href
  {http://adsabs.harvard.edu/abs/2010ApJ...719..722S} {719, 722}

\bibitem[\protect\citeauthoryear{{Shapiro} \& {Teukolsky}}{{Shapiro} \&
  {Teukolsky}}{1983}]{st83}
{Shapiro} S.~L.,  {Teukolsky} S.~A.,  1983, {Black holes, white dwarfs, and
  neutron stars: The physics of compact objects}.
New York, Wiley-Interscience

\bibitem[\protect\citeauthoryear{{Shen}, {Toki}, {Oyamatsu}  \&
  {Sumiyoshi}}{{Shen} et~al.}{1998}]{HSHEN-4}
{Shen} H.,  {Toki} H.,  {Oyamatsu} K.,   {Sumiyoshi} K.,  1998, \mn@doi
  [Progress of Theoretical Physics] {10.1143/PTP.100.1013}, \href
  {http://adsabs.harvard.edu/abs/1998PThPh.100.1013S} {100, 1013}

\bibitem[\protect\citeauthoryear{{Spitkovsky}}{{Spitkovsky}}{2006}]{spitkovsky06}
{Spitkovsky} A.,  2006, \mn@doi [\apjl] {10.1086/507518}, \href
  {http://adsabs.harvard.edu/abs/2006ApJ...648L..51S} {648, L51}

\bibitem[\protect\citeauthoryear{{Sugahara} \& {Toki}}{{Sugahara} \&
  {Toki}}{1994}]{HSHEN-5}
{Sugahara} Y.,  {Toki} H.,  1994, \mn@doi [Nuclear Physics A]
  {10.1016/0375-9474(94)90923-7}, \href
  {http://adsabs.harvard.edu/abs/1994NuPhA.579..557S} {579, 557}

\bibitem[\protect\citeauthoryear{{Suleimanov}, {Poutanen}, {Revnivtsev}  \&
  {Werner}}{{Suleimanov} et~al.}{2011}]{SLM}
{Suleimanov} V.,  {Poutanen} J.,  {Revnivtsev} M.,   {Werner} K.,  2011,
  \mn@doi [\apj] {10.1088/0004-637X/742/2/122}, \href
  {http://adsabs.harvard.edu/abs/2011ApJ...742..122S} {742, 122}

\bibitem[\protect\citeauthoryear{{Tauris} \& {Manchester}}{{Tauris} \&
  {Manchester}}{1998}]{tm98}
{Tauris} T.~M.,  {Manchester} R.~N.,  1998, \mn@doi [\mnras]
  {10.1046/j.1365-8711.1998.01369.x}, \href
  {http://adsabs.harvard.edu/abs/1998MNRAS.298..625T} {298, 625}

\bibitem[\protect\citeauthoryear{{Tchekhovskoy}, {Spitkovsky}  \&
  {Li}}{{Tchekhovskoy} et~al.}{2013}]{tche13}
{Tchekhovskoy} A.,  {Spitkovsky} A.,   {Li} J.~G.,  2013, \mn@doi [\mnras]
  {10.1093/mnrasl/slt076}, \href
  {http://adsabs.harvard.edu/abs/2013MNRAS.435L...1T} {435, L1}

\bibitem[\protect\citeauthoryear{{Thorsett} \& {Chakrabarty}}{{Thorsett} \&
  {Chakrabarty}}{1999}]{thor99}
{Thorsett} S.~E.,  {Chakrabarty} D.,  1999, \mn@doi [\apj] {10.1086/306742},
  \href {http://adsabs.harvard.edu/abs/1999ApJ...512..288T} {512, 288}

\bibitem[\protect\citeauthoryear{{Tiengo} et~al.,}{{Tiengo}
  et~al.}{2013}]{tien13}
{Tiengo} A.,  et~al., 2013, \mn@doi [\nat] {10.1038/nature12386}, \href
  {http://adsabs.harvard.edu/abs/2013Natur.500..312T} {500, 312}

\bibitem[\protect\citeauthoryear{{Tsang} \& {Gourgouliatos}}{{Tsang} \&
  {Gourgouliatos}}{2013}]{tsang13}
{Tsang} D.,  {Gourgouliatos} K.~N.,  2013, \mn@doi [\apjl]
  {10.1088/2041-8205/773/1/L17}, \href
  {http://adsabs.harvard.edu/abs/2013ApJ...773L..17T} {773, L17}

\bibitem[\protect\citeauthoryear{{Typel}, {R{\"o}pke}, {Kl{\"a}hn}, {Blaschke}
  \& {Wolter}}{{Typel} et~al.}{2010}]{BHB-3}
{Typel} S.,  {R{\"o}pke} G.,  {Kl{\"a}hn} T.,  {Blaschke} D.,   {Wolter} H.~H.,
   2010, \mn@doi [\prc] {10.1103/PhysRevC.81.015803}, \href
  {http://adsabs.harvard.edu/abs/2010PhRvC..81a5803T} {81, 015803}

\bibitem[\protect\citeauthoryear{{Urama}, {Link}  \& {Weisberg}}{{Urama}
  et~al.}{2006}]{urama06}
{Urama} J.~O.,  {Link} B.,   {Weisberg} J.~M.,  2006, \mn@doi [\mnras]
  {10.1111/j.1745-3933.2006.00192.x}, \href
  {http://adsabs.harvard.edu/abs/2006MNRAS.370L..76U} {370, L76}

\bibitem[\protect\citeauthoryear{{Vigan{\`o}}, {Rea}, {Pons}, {Perna},
  {Aguilera}  \& {Miralles}}{{Vigan{\`o}} et~al.}{2013}]{vigano13}
{Vigan{\`o}} D.,  {Rea} N.,  {Pons} J.~A.,  {Perna} R.,  {Aguilera} D.~N.,
  {Miralles} J.~A.,  2013, \mn@doi [\mnras] {10.1093/mnras/stt1008}, \href
  {http://adsabs.harvard.edu/abs/2013MNRAS.434..123V} {434, 123}

\bibitem[\protect\citeauthoryear{{Vrane{\v s}evi{\'c}} \& {Melrose}}{{Vrane{\v
  s}evi{\'c}} \& {Melrose}}{2011}]{vm11}
{Vrane{\v s}evi{\'c}} N.,  {Melrose} D.~B.,  2011, \mn@doi [\mnras]
  {10.1111/j.1365-2966.2010.17612.x}, \href
  {http://adsabs.harvard.edu/abs/2011MNRAS.410.2363V} {410, 2363}

\bibitem[\protect\citeauthoryear{{Weltevrede} \& {Johnston}}{{Weltevrede} \&
  {Johnston}}{2008}]{welt08}
{Weltevrede} P.,  {Johnston} S.,  2008, \mn@doi [\mnras]
  {10.1111/j.1365-2966.2008.13382.x}, \href
  {http://adsabs.harvard.edu/abs/2008MNRAS.387.1755W} {387, 1755}

\bibitem[\protect\citeauthoryear{{Wiringa}, {Fiks}  \& {Fabrocini}}{{Wiringa}
  et~al.}{1988}]{WFF}
{Wiringa} R.~B.,  {Fiks} V.,   {Fabrocini} A.,  1988, \mn@doi [\prc]
  {10.1103/PhysRevC.38.1010}, \href
  {http://adsabs.harvard.edu/abs/1988PhRvC..38.1010W} {38, 1010}

\bibitem[\protect\citeauthoryear{{Xu} \& {Qiao}}{{Xu} \& {Qiao}}{2001}]{xu01}
{Xu} R.~X.,  {Qiao} G.~J.,  2001, \mn@doi [\apjl] {10.1086/324381}, \href
  {http://adsabs.harvard.edu/abs/2001ApJ...561L..85X} {561, L85}

\bibitem[\protect\citeauthoryear{{Xu} \& {Wu}}{{Xu} \& {Wu}}{1991}]{xuwu91}
{Xu} W.,  {Wu} X.,  1991, \mn@doi [\apj] {10.1086/170612}, \href
  {http://adsabs.harvard.edu/abs/1991ApJ...380..550X} {380, 550}

\bibitem[\protect\citeauthoryear{{Young}, {Chan}, {Burman}  \& {Blair}}{{Young}
  et~al.}{2010}]{young10}
{Young} M.~D.~T.,  {Chan} L.~S.,  {Burman} R.~R.,   {Blair} D.~G.,  2010,
  \mn@doi [\mnras] {10.1111/j.1365-2966.2009.15972.x}, \href
  {http://adsabs.harvard.edu/abs/2010MNRAS.402.1317Y} {402, 1317}

\bibitem[\protect\citeauthoryear{{Zhang} \& {Xie}}{{Zhang} \&
  {Xie}}{2012}]{zhang12}
{Zhang} S.-N.,  {Xie} Y.,  2012, \mn@doi [\apj] {10.1088/0004-637X/761/2/102},
  \href {http://adsabs.harvard.edu/abs/2012ApJ...761..102Z} {761, 102}

\bibitem[\protect\citeauthoryear{{Zhang}, {Jiang}  \& {Mei}}{{Zhang}
  et~al.}{2003}]{zjm03}
{Zhang} L.,  {Jiang} Z.-J.,   {Mei} D.-C.,  2003, \mn@doi [\pasj]
  {10.1093/pasj/55.2.461}, \href
  {http://adsabs.harvard.edu/abs/2003PASJ...55..461Z} {55, 461}

\bibitem[\protect\citeauthoryear{{van Kerkwijk}, {Breton}  \& {Kulkarni}}{{van
  Kerkwijk} et~al.}{2011}]{Kerk}
{van Kerkwijk} M.~H.,  {Breton} R.~P.,   {Kulkarni} S.~R.,  2011, \mn@doi
  [\apj] {10.1088/0004-637X/728/2/95}, \href
  {http://adsabs.harvard.edu/abs/2011ApJ...728...95V} {728, 95}

\makeatother
\end{thebibliography}
\input{magnetic_field_refined.bbl}



\appendix

\section{Calculations of the parameters of a neutron star}
\label{sect:eos_calc}
Let us consider the basic moments of calculations of NS parameters. Note that a simple estimation of the Kepler frequency
$\Omega_{k}\approx \sqrt{GM/R^3}$ shows that for $M=1.4M_{\odot}$ its value lies between
7.4 and 16 kHz for $9<R<15$ km. The frequencies of rotation
for the considered pulsars are much lower, and therefore the slow-rotation approximation can be used for solving the general relativity
equations. The stellar configurations are assumed to be spherical.
The space-time metric with only first-order rotational terms with
respect to the stellar angular velocity ${\Omega}$ can be written as

\begin{table*}
\centering
\caption{The parameters of neutron stars for various equations of state discussed in the paper. Neutron stars radii and moments of inertia
for the $M=1.4M_{\odot}$ and $M=1.49M_{\odot}$ are shown, as well as the
maximal mass limit for corresponding equation of state. The means $\langle \cdot \rangle$ and standard deviations $\sigma[\cdot]$ of the
correction $\Delta_{\rm B}^{\rm (eos)}$ are calculated adopting the mass distribution (\ref{eq:mass_distrib}) and anisotropic
obliquities (\ref{eq:alpha_distrib}). The horizontal line separate equations of state based on the DBHF approach and many-body calculations (above) from ones based on relativistic mean field theory (below).}
\label{tab:NSC}
\begin{centering}
\begin{tabular}{lccccccc}
  \hline
       &                         & \multicolumn{2}{c}{$M = 1.4 M_{\sun}$} & \multicolumn{2}{c}{$M = 1.49 M_{\sun}$} \\
   EOS & $M_{\rm max}/M_{\sun}$  & $R$, km  & $I_{45}$                  & $R$, km & $I_{45}$ & $\langle \Delta_{\rm B}^{\rm (eos)}\rangle$ & $\sigma[\Delta_{\rm B}^{\rm (eos)}]$ \\
  \hline
  SLy4              & 2.05 & 11.69   & 1.36 & 11.62   & 1.49 & -0.279 &  0.080 \\
  WFF2              & 2.21 & 11.04   & 1.28 & 11.03   & 1.39 & -0.224 &  0.075 \\
  BSK20             & 2.17 & 11.74   & 1.38 & 11.71   & 1.53 & -0.283 &  0.075 \\
  BSK21             & 2.28 & 12.58   & 1.57 & 12.58   & 1.73 & -0.349 &  0.072 \\
  AP3               & 2.38 & 12.06   & 1.49 & 12.06   & 1.63 & -0.308 &  0.071 \\
  AP4               & 2.19 & 11.36   & 1.33 & 11.29   & 1.44 & -0.250 &  0.074 \\
  SK16              & 2.19 & 12.47   & 1.53 & 12.44   & 1.69 & -0.338 &  0.074 \\
  MPA1              & 2.49 & 12.39   & 1.57 & 12.41   & 1.70 & -0.334 &  0.069 \\
  ENG               & 2.23 & 11.85   & 1.42 & 11.86   & 1.54 & -0.293 &  0.072 \\
 \hline
  GM1               & 2.39 & 14.19   & 1.85 & 14.17   & 2.03 & -0.467 &  0.074 \\
  GM1Y5             & 2.12 & 13.78   & 1.86 & 13.78   & 2.03 & -0.430 &  0.070 \\
  GM1Y6             & 2.30 & 13.78   & 1.86 & 13.78   & 2.03 & -0.430 &  0.069 \\
  DDH$\Delta$       & 2.16 & 12.65   & 1.63 & 12.65   & 1.78 & -0.348 &  0.070 \\
  DDH$\Delta Y$4    & 2.05 & 12.65   & 1.63 & 12.65   & 1.78 & -0.348 &  0.071 \\
  BHB$\Lambda\phi$  & 2.10 & 12.95   & 1.74 & 12.98   & 1.90 & -0.367 &  0.069 \\
  BHB$\Lambda$      & 1.95 & 12.96   & 1.74 & 12.98   & 1.90 & -0.366 &  0.069 \\
  MS1               & 2.77 & 14.83   & 2.05 & 14.85   & 2.25 & -0.507 &  0.070 \\
  MS1b              & 2.78 & 14.52   & 2.00 & 14.55   & 2.20 & -0.487 &  0.068 \\
  GNH3              & 1.97 & 14.18   & 1.80 & 14.00   & 1.94 & -0.462 &  0.087 \\
  H4                & 2.03 & 12.87   & 1.84 & 12.99   & 1.98 & -0.428 &  0.074 \\
  ALF2              & 2.09 & 13.17   & 1.74 & 13.20   & 1.91 & -0.388 &  0.069 \\
  HShen+QB139       & 2.30 & 13.16   & 1.91 & 13.27   & 2.07 & -0.377 &  0.062 \\
  \hline
\end{tabular}
\end{centering}
\end{table*}

\begin{equation}
	\begin{aligned}
    ds^2= & - e^{2\psi(r)}c^{2}dt^2 + e^{2\lambda(r)}dr^2 + \\
		  & r^2\left(d\theta^2+\sin^{2}\theta\left(d\phi^{2}-2(\Omega-\omega(r,\theta))d\phi dt\right)\right).
    \end{aligned}
\end{equation}
Here $\psi$ and $\lambda$ are the functions of radial coordinate only.
The value $\hat{\omega}=\Omega-\omega(r,\theta)$ is nothing else
than the angular velocity of zero-angular-momentum observer. The
Einstein equations are
\begin{equation}
    R_{\mu\nu}-\frac{1}{2}g_{\mu\nu}R=\frac{8\upi G}{c^4}T_{\mu\nu}.
\end{equation}
Here $R_{\mu\nu}$ is Ricci tensor for the metric $g_{\mu\nu}$, $R$ is
the scalar curvature and $T_{\mu\nu}$ is the energy-momentum tensor for
stellar matter. For the case of spherical symmetry
\begin{equation}
    T^{\mu}_{\nu}=(\rho c^2+p)u^{\mu}u_{\nu}-p\delta^{\mu}_{\nu},
\end{equation}
where $u^{\mu}$ is the matter 4-velocity. For axial symmetry and
uniform rotation we have for $u^{\mu}$ the following relation:
\begin{equation}
    u^{\mu}=u^{t}(1,0,0,\Omega).
\end{equation}
Here $\rho$ and $p$ are density and pressure of matter
respectively. Keeping only first order terms with respect to
$\hat{\omega}$ one can write the components of the field equations as
\begin{equation}
 \label{TOV1-1}
 \frac{1}{r^2}\frac{d}{dr}\left[ r\left(1-e^{-2\lambda }\right)\right] = 8\upi G \rho \/c^{2},
\end{equation}
\begin{equation}
 \label{TOV2-1}
 \frac{1}{r} \left[2e^{-2\lambda}\frac{d\psi}{dr}-\frac{1}{r}\left(1-e^{-2\lambda}\right)\right] = 8\upi p G/c^4.
\end{equation}
Taking into account the hydrostatic equilibrium condition
\begin{equation}
    \label{hydro-1}
    \frac{dp}{dr}=-(\rho c^2+p)\frac{d\psi}{dr}
\end{equation}
and definition of gravitational mass $m(r)$ according to the relationship
\begin{equation}
    e^{-2\lambda(r)}=1-\frac{2G m(r)}{c^2 r}
\end{equation}
it can be concluded that the equations above are nothing other than
the ordinary Tolmen-Oppenheimer-Volkoff equations.

For $\omega(r,\theta)$ we have the equation
\begin{equation}
    \label{TOV4-1}
    \begin{aligned}
    \frac{e^{\psi-\lambda}}{r^4} \partial_{r}\left[e^{-(\psi +
    \lambda)} r^4 \partial_{r}{\omega} \right]  +
    \frac{1}{r^2\sin^3\theta}
    \partial_{\theta}\left[\sin^3\theta\partial_{\theta}\omega
    \right]= \\
	\frac{16\upi G}{c^4}(\rho c^2 + p){\omega}.
    \end{aligned}
\end{equation}
For asymptotically flat space-time, the angular velocity is a
function of the radial coordinate only and therefore equation
(\ref{TOV4-1}) can be rewritten as
\begin{equation}
    \label{OR}
    \frac{e^{\psi-\lambda}}{r^4} \frac{d}{dr}\left[e^{-(\psi+
    \lambda)}r^4 \frac{d{\omega}}{dr} \right] = \frac{16\upi G}{c^4}
    (\rho c^2 + p){\omega}.
\end{equation}
At $r\rightarrow\infty$, the following condition on $\omega(r)$
should be satisfied:
\begin{equation}
    \lim_{r\to \infty}{\omega(r)}={\Omega}.
\end{equation}
Finally, the regularity condition at the center of a star requires
that
\begin{equation}
    \frac{d{\omega}(0)}{dr}= 0.
\end{equation}
The moment of inertia is defined via the angular momentum  $J$
according to the relationship
\begin{equation}
    I\equiv \frac{J}{\Omega}.
\end{equation}
The Angular moment is
\begin{equation}
    J=\int T^{\mu}_{\nu}\xi^{\nu}_{(\phi)}\sqrt{-g} d^{3}x.
\end{equation}
Here $\xi^{\nu}_{(\phi)}$ is the Killing vector in the azimuthal
direction. Keeping only terms of first order in ${\omega}$ the moment of inertia
can be evaluated as
\begin{equation}\label{Inertial}
    I \approx \frac{8\upi}{3} \int_{0}^{R}(\rho + p/c^2)e^{\lambda - \psi} r^4 \frac{{\omega}}{\Omega} dr .
\end{equation}
Therefore the moment of inertia in the slow-rotation approximation is
independent from the angular velocity.


\bsp    
\label{lastpage}
\end{document}